\documentclass[12pt]{iopart}
\usepackage{color}
\usepackage{graphicx}
\usepackage{iopams}
\usepackage{dsfont}
\usepackage{enumitem}
\usepackage{arydshln}

\def\arctanh{\mathrm{arctanh}}
\def\sinh{\mathrm{sinh}}
\usepackage[colorlinks=true,linkcolor=blue,citecolor=blue]{hyperref}

\begin{document}
\title{From classical to quantum non-equilibrium dynamics of Rydberg excitations in optical lattices}
\author{Marco Mattioli$^{1,2}$, Alexander W. Gl{\"a}tzle$^{1,2}$ and Wolfgang Lechner$^{1,2}$}

\address{$^{1}$Institute for Quantum Optics and Quantum Information of the Austrian Academy of Sciences, A-6020 Innsbruck, Austria}
\address{$^{2}$Institute for Theoretical Physics, University of Innsbruck, A-6020 Innsbruck, Austria}

\ead{marco.mattioli@uibk.ac.at}
\begin{abstract}  
The glass phase and its quantum analog are prominent challenges of current non-equilibrium statistical mechanics and condensed matter physics. As a model system to study the transition from classical to quantum glassy dynamics, we propose a setup of laser driven three-level atoms trapped in an optical lattice. Tuning the strength of the laser driving to the intermediate level allows one to study the transition from a classical Kinetically Constrained Model to the coherent regime.
For strong driving, Rydberg excitations evolve analogously to defects in the One-Spin Facilitated Model, a minimal model known to exhibit glassy dynamics. In our setup, the constraints result from the interplay between Rydberg interactions and the laser detuning from the Rydberg state. The emerging heterogeneous relaxation timescales are tuneable over several orders of magnitudes. In the opposite limit of weak driving of the intermediate level, we find an effective cluster model which describes the dynamics in a reduced subspace of the allowed number and positions of Rydberg excitations. This subspace is uniquely determined by the initial state and is characterized by a fixed number of clusters of Rydberg excitations. In addition, we investigate the influence of random fields on the classical relaxation. We find that the glassy dynamics can relax faster in the presence of weak random fields.
\end{abstract}

\maketitle

\section{Introduction}

When disordered materials are quenched to low temperatures, the dynamics can become non-ergodic and the system may fall out of equilibrium~\cite{book_binder,book_goetze,rev:glass_deb, rev:glass_bir,rev:glass_chand}. Understanding this phenomenon, known as the glass transition, is one of the open questions in current statistical mechanics. While a complete theory which satisfactorily explains all features of glass forming materials is missing, spin models have been proposed to understand particular aspects of the glass problem. For example, static properties of glasses are captured by Ising models with random interactions, the so-called spin glasses~\cite{PARISI,kurtbinder,EASM}. These models have recently gained renewed interest in the context of both quantum adiabatic optimization~\cite{farhi,troyeraqo} and, in presence of external random fields, of the many-body localization phase transition~\cite{anderson,altsh,rev:mbl,huse}. Spin models which study dynamical properties of glasses are so-called Kinetically Constrained Models~\cite{rev:fsm} (KCMs). In these KCMs, separation of relaxation timescales as well as space-time dynamical heterogeneities~\cite{rev:glass_chand, tanaka} result from transitions (spin flips) which are allowed or forbidden depending on the state of neighboring spins. 
Quantum effects in KCMs have been recently studied in Refs.~\cite{QUANTUMGLASS,lesanov_fsm,olmos,lesanov_kc,marcuzzi} in ultracold Rydberg atoms.
Thanks to their strong and long-range interactions, as well as to their long lifetimes~\cite{book:rydberg,rev:rydsaff,rev:rydgall}, Rydberg atoms offer a promising framework to study quantum spin models in- and out-of-equilibrium~\cite{ates, weimer, lukin, lesanov,lesanov1, ji, entropy,honing,fleisch,petr,evers,glaetzle}. With the considerable recent experimental development in atomic, molecular and optical physics, Rydberg atoms can in fact be trapped, laser-excited and detected with single particle resolution in state-of-the-art experiments~\cite{RYDBERGCYRSTAL,gunter, hofmann, robert,weide,morsch, valado,hott}.

\begin{figure}[htb] 
\centerline{\includegraphics[width=0.75\textwidth]{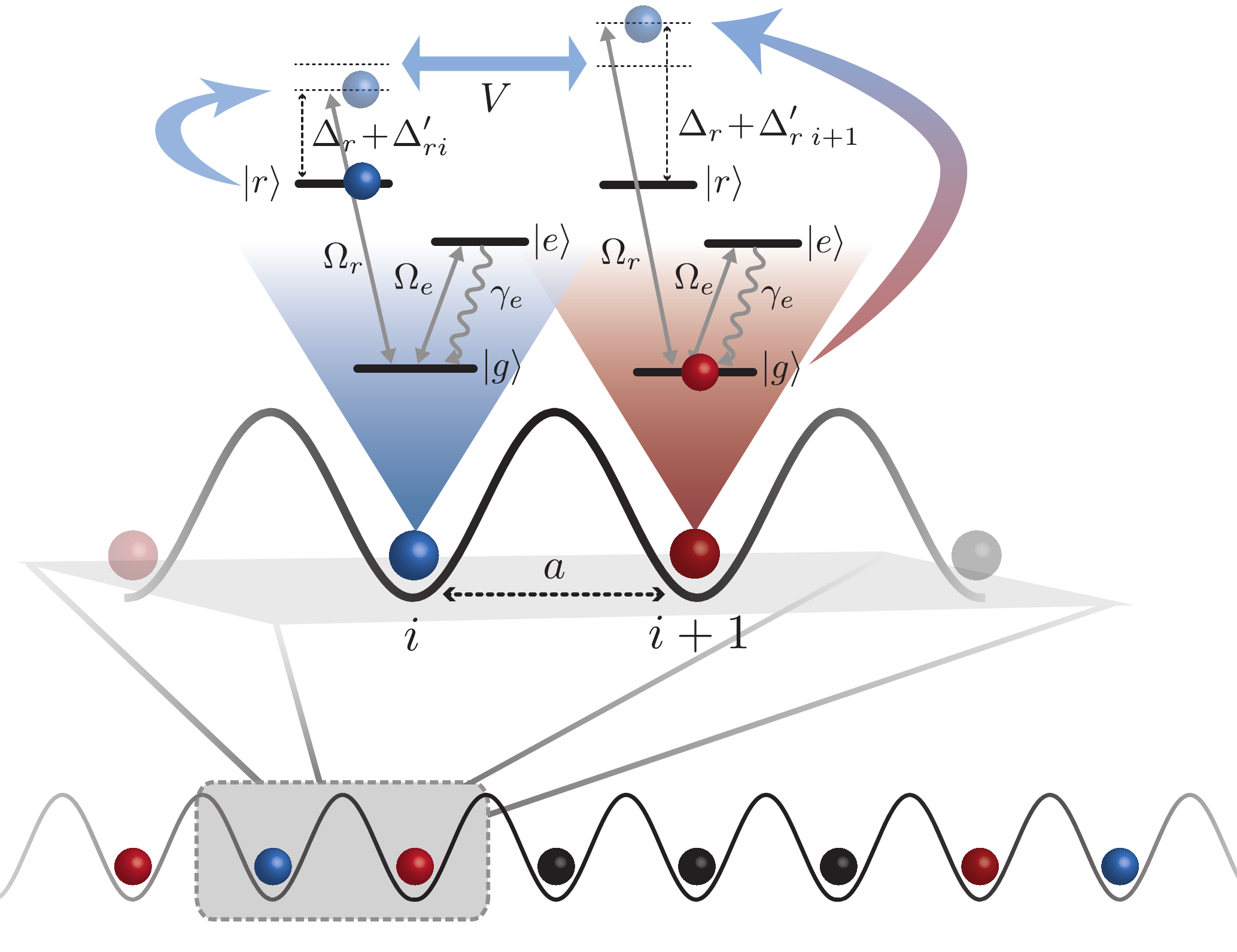}}
\caption{The setup we have in mind consists of three-level atoms, each one schematized with a ground-state $|g\rangle$,  a Rydberg state $|r\rangle$ and an intermediate excited state $|e\rangle$. Atoms are coherently laser-driven with Rabi frequencies $\Omega_e$ ($\Omega_r$) between $|g\rangle$-$|e\rangle$ ($|g\rangle$-$|r\rangle$) and $\gamma_e$ is the decay rate due to spontaneous emission from $|e\rangle$. 
The anti-blockade condition $V = \Delta_r$, i.e. when the laser detuning $\Delta_r$ with respect to the $|g\rangle$-$|r\rangle$ atomic transition is equal to the interactions between neighbouring Rydberg-excited atoms $V$ (see Subsec.~\ref{many_body}), ensures that whenever an atom is in $|r\rangle$ (\textit{defect}, blue sphere), an atom next to it, initially in $|g\rangle$ (\textit{facilitated} atom, red sphere), will also be favoured to be excited to $|r\rangle$. An atom in $|g\rangle$ far from a defect (\textit{non-facilitated} atom, black sphere) is blocked, because its  $|g\rangle$-$|r\rangle$ transition is far-off-resonant (see Subsec.~\ref{subsec:withoutd}). The case with additional site-dependent random detunings $\Delta^{\prime}_{ri}$ is studied in Subsec.~\ref{subsec:extdis}.}
\label{fig:scheme}
\end{figure}

In this work we propose a model system based on V-shaped three-level atoms, that allows one to study the transition between a classical dissipation-induced glass model and a fully coherent quantum system. Atoms are resonantly driven from the electronic ground-state $|g\rangle$ to a short-lived low-lying excited state $|e\rangle$ and far-off-resonantly driven to a metastable Rydberg state $|r\rangle$ (see Fig.~\ref{fig:scheme}). Tuning the driving strength $\Omega_e$ of the $|g\rangle \leftrightarrow |e\rangle$ transition, allows one to study the crossover from classical to quantum dynamics of Rydberg excitations. 
For large $\Omega_e$, the relaxation of Rydberg excitations resembles the one of defects in a classical KCM, the One-Spin Facilitated Model~\cite{rev:fsm,lesanov_fsm} (1-SFM). Here, kinetic constraints are a consequence of the interplay between strong interactions of Rydberg-excited atoms~\cite{dipblock} and the laser detuning from $|r\rangle$. In contrast, for vanishing $\Omega_e$, coherences of the $|g\rangle \leftrightarrow |r\rangle$ transition dominate the dynamics within the lifetime of the Rydberg states~\cite{RYDBERGCYRSTAL}. In the case of large laser detuning from the Rydberg state the system resembles a transverse Ising model for a particular choice of the detuning~\cite{calabrese1,calabrese2}. We present an effective simple model to describe the evolution in this regime which is based on a fixed number of Rydberg clusters with variable number of Rydberg excitations.

In addition to the study of the transition from a glass model to a fully coherent integrable model, the proposed framework opens up the possibility to investigate the interplay between internal dynamical constraints and external disorder arising from random fields. In our system, static random fields can be implemented with site-dependent random detunings from the Rydberg state. Our results indicate that the glassy relaxation of Rydberg excitations toward the steady state can be accelerated by the presence of weak random fields.

The remainder of the paper is organized as follows: In Sec.~\ref{sec:model} the level-scheme and the model are introduced; in Sec.~\ref{subsec:fsm} we give a brief overview of SFMs. Numerical results from quantum trajectory simulations are presented in Sec.~\ref{sec:num_res}. In Subsec.~\ref{subsec:sa_re} we study the coherent regime and the transition to the rate equation limit and in Subsec.~\ref{subsec:extdis} we analyze the case of finite random detunings from the Rydberg state. The validity of the assumptions and approximations of Sec.~\ref{sec:num_res} are discussed in Sec.~\ref{sec:limit}. Finally, in Sec.~\ref{sec:conc} we summarize our conclusions.

\section{Model} \label{sec:model}

We consider $N$ atoms individually trapped in a one-dimensional ($1d$) large spacing optical lattice~\cite{RYDBERGCYRSTAL,nelson,kuzmich,Browaeys,Schlosser} or in a magnetic trap array~\cite{Hannaford, Spreeuw}. Atomic motion due to tunnelling between adjacent sites is negligible within experimental timescales.
Each atom is approximated as a three-level system in the V-configuration, as schematically depicted in Fig.~\ref{fig:scheme}: the electronic ground-state $|g\rangle$ is laser driven to (i) a Rydberg state $|r \rangle$ with a small Rabi frequency $\Omega_r$, and (ii) to an excited state $|e \rangle$ with a large Rabi frequency $\Omega_e$. The intermediate low-lying excited state $|e \rangle$ is rapidly decaying to $|g\rangle$ and its short lifetime is given by the inverse of the decay rate $\gamma_e$. Two atoms in their Rydberg states interact via dipole-dipole $V(x) = V/x^3$ or van der Waals $V(x) = V/x^6$ potentials, depending on the presence or absence of an applied external field~\cite{rev:rydsaff}, respectively. We approximate interactions between Rydberg-excited atoms to be nearest-neighbor, i.e. $V(1) = V$ and $V(x \geqslant 2) = 0$. Results under this assumption are in qualitative good agreement with simulations of the full model including long-range interactions. Furthermore, we neglect processes of spontaneous emission from the Rydberg state, setting the associated decay rate $\gamma_r$ to zero. Restrictions on the validity of these approximation are discussed in detail in Sec.~\ref{sec:limit}.

The Hamiltonian in the rotating frame (with $\hbar = 1$) is

\begin{eqnarray}  \label{newham}
H = \sum_{i=1}^N \Biggl \{&  \frac{\Omega_e}{2} \bigg(|g \rangle \langle e |_i + \mathrm{H.c.}\bigg) + \frac{\Omega_r}{2} \bigg(|g \rangle \langle r |_i + \mathrm{H.c.}\bigg) + \nonumber \\ &- \bigg(\Delta_r+\Delta_{ri}^{\prime}\bigg) |r \rangle \langle r |_i \Biggr \} +  
 V \sum_{i=1}^{N-1} \biggl \{ |r \rangle \langle r |_i \otimes |r \rangle \langle r |_{i+1}  \biggr \}.
\end{eqnarray}  
Here, the sum runs over all $N$ lattice sites, while $\Delta_r$ and $\Delta^{\prime}_{ri}$ are the homogeneous and site-dependent random laser detunings from $|r\rangle$, respectively. 
The random detuning can, for example, be induced by a spatially inhomogeneous ac-Stark shift acting on the Rydberg- or ground-state using either a random light pattern or differential light-shifts between atoms in different traps, in close analogy to experiments performed in the context of Anderson localization~\cite{aspect,roati}.

\subsection{A single-atom in the V-configuration} \label{subsec:sp}

Even though a single three-level systems in the V-configuration has been extensively studied in literature, both theoretically~\cite{dal_qjumps,pz,zol_qjumps,schenzle,tbtd,cohen_dal,rev} and experimentally~\cite{dehmelt,wineland}, let us briefly discuss its properties. For a more complete and detailed discussion, we refer the reader to~\ref{appendix:st}.

The single-atom Hamiltonian of the three-level scheme in Eq.~(\ref{newham}), where the Rydberg-state is far-off-resonantly laser driven ($\Delta_r \neq 0$) and no random detuning is present ($\Delta^{\prime}_{ri} = 0$, $\forall i \in [1,N]$), is
\begin{equation} \label{ham_sa}
H = \frac{\Omega_e}{2} \big( |g \rangle \langle e| + \mathrm{H.c.} \big) + \frac{\Omega_r}{2} \big( |g \rangle \langle r| + \mathrm{H.c.}  \big) -\Delta_r |r\rangle \langle r|.
\end{equation}
In the Born-Markov approximation, the master equation describing the dissipative evolution of the reduced system density matrix $\rho(t)$ is in the Lindblad form~\cite{book_gardiner_zoller}
\begin{equation} \label{me}
\dot{\rho}(t) = - i [H,\rho(t)] + \frac{\gamma_e}{2}  \big( 2c\rho(t) c^{\dagger} - c^{\dagger} c \rho(t) - \rho(t) c^{\dagger} c \big),
\end{equation}
where $c = |g \rangle \langle e|$ is the quantum jump operator. 
 Assuming $\Omega_r \ll \Omega_e^2/\gamma_e$ and $\Delta_r \gg \Omega_e,\Omega_r, \gamma_e$ (the reason behind this choice is explained in~\ref{appendix:st}), we imagine to measure photons emitted by our three-level atom.  A typical photon counting sequence registered by a photodetector would then look like an alternation of time intervals with many photons detected (\textit{bright} periods, which we label with $0$) and no photons detected (\textit{dark} periods, which we label with $1$).
The transition probability (rate) in going from $0$ to $1$ is given by

\begin{equation} \label{sprateup}
\Gamma^{0 \rightarrow 1} (\Delta_r) = \frac{\gamma_e \Omega_e^2\Omega_r^2(\gamma_e^2 + 4\Delta_r^2)}{(\gamma_e^2 + 2\Omega_e^2)[16 \Delta_r^4 + 4 \Delta_r^2 (\gamma_e^2 - 2\Omega_e^2) + \Omega_e^4]},
\end{equation}
whereas the transition probability from $1$ to $0$ is

\begin{equation} \label{sprate}
\Gamma^{1 \rightarrow 0} (\Delta_r) =  \frac{\gamma_e \Omega_e^2 \Omega_r^2}{16 \Delta_r^4 + 4 \Delta_r^2 (\gamma_e^2 - 2\Omega_e^2) + \Omega_e^4}.
\end{equation}
The analytical derivation of these formulas and the physical mechanism behind these transitions, known as \textit{quantum jumps}, are given in~\ref{appendix:st}.

\subsection{Many-body model} \label{many_body}
In this Subsection we study the many-body model ($\Delta_r, V \neq 0$) with no random detunings. In analogy to the single-atom case [Eq.~(\ref{me})], the many-body master equation

\begin{equation} \label{ffme}
\dot{\rho}(t) = -i[H, \rho(t)] + \frac{\gamma_e}{2} \sum_{i =1}^{N} (2 c_{i} \rho(t) c^{\dagger}_{i} - c^{\dagger}_{i} c_i \rho(t) - \rho(t) c^{\dagger} _{i} c_{i}),
\end{equation}
with $c_i = |g\rangle \langle e|_i$ the quantum jump operator at site $i$, describes the driven-dissipative dynamics of the many-body system.
Furthermore, let us assume: (i) $\Omega_r \ll \Omega_e^2/\gamma_e$, as in Subsec.~\ref{subsec:sp}, for well defined \textit{quantum jumps} to occur and (ii) $\Delta_r \gg \Omega_e, \Omega_r,\gamma_e$ and (iii) $V = \Delta_r$, which we call the \textit{anti-blockade} condition.
The many-body transition rates, corresponding to the single-particle rates in Eqs.~(\ref{sprateup}) and~(\ref{sprate}), can be determined as follows. In a $1d$ system, due to the assumption $V(x \geqslant 2) = 0$, every  process at a certain site $i$ can be influenced at most by its two nearest-neighbors, $i\pm1$. The many-body rates can be calculated from Eqs.~(\ref{sprateup}) and~(\ref{sprate}) with $\Delta^{\star}_r = \Delta_r - \ell V$ the effective detuning and $\ell \in [0,2]$ the number of neighbors in the Rydberg state. Due to the \textit{anti-blockade} condition, there are three possible configurations:  $\Delta^{\star}_r = \Delta_r$ if there are no neighbors in the Rydberg state ($\ell=0$), $\Delta^{\star}_r = 0$ if there is only one neighbor in the Rydberg state ($\ell=1$) and $\Delta^{\star}_r = -\Delta_r$ if both neighbors are in the Rydberg state ($\ell=2$). Since in Eqs.~(\ref{sprateup}) and~(\ref{sprate}), $\Delta_r$ appears only in even powers, both off-resonant rates $\ell=0,2$ are  equally suppressed with respect to the resonant rate $\ell=1$, as highlighted in Tab.~\ref{tab:rates_off}. 

\begin{table}[h!]
\centering
 \begin{tabular}{||c | c |  c | c ||} 
 \hline
 & $\Delta_r^{\star} = 0$  &  $\Delta_r^{\star} = \pm3\gamma_e$    & $\Delta_r^{\star} = \pm10\gamma_e$ \\ [0.5ex] 
 \hline
  $\Gamma^{0 \rightarrow 1}(\Delta_r^{\star})/\gamma_e$ &   $3.00 \cdot 10^{-4}$  &   $8.80 \cdot 10^{-6}$   &   $7.54 \cdot 10^{-7}$\\  [1ex] 
 \hline
  $\Gamma^{1 \rightarrow 0}(\Delta_r^{\star})/\gamma_e$ &   $9.00 \cdot 10^{-4}$  &  $7.14 \cdot 10^{-7}$  &  $5.64 \cdot 10^{-9}$ \\  [1ex] 
  \hline
 \end{tabular}
 \caption{Calculation of the creation [$\Gamma^{0 \rightarrow 1}(\Delta_r^{\star})$] and destruction [$\Gamma^{1 \rightarrow 0}(\Delta_r^{\star})$] rates  of a Rydberg excitation (in units of $\gamma_e$), from Eqs.~(\ref{sprateup}) and~(\ref{sprate}), for $\Delta_r^{\star} = 0$ (resonant rates) and $\Delta_r^{\star} = \pm 3\gamma_e, \pm 10\gamma_e$ (far-off-resonant rates). Other parameters are: $\Omega_e = \gamma_e$ and $\Omega_r = 0.03\gamma_e$. The difference between resonant and far-off-resonant rates is of several order of magnitudes and increases with $|\Delta_r^{\star}|$. } \label{tab:rates_off}
\end{table}

In absence of conditions (i), (ii) and (iii), the strong transition $|g\rangle \leftrightarrow |e\rangle$ might significantly increase the probability $\Gamma^{0(1)\rightarrow 1(0)}(\pm \Delta_r)$ of (originally) suppressed processes of excitation (deexcitation) to (from) $|r\rangle$ to take place. Since rates differ by several order of magnitudes (see Tab.~\ref{tab:rates_off}), the dynamics shows a large separation of timescales (cfr. Sec.~\ref{sec:num_res}).

Allowed ($\ell=1$) and suppressed ($\ell=0,2$) processes are analogous to those who characterize Spin Facilitated Models, as discussed in Sec.~\ref{subsec:fsm}.

\section{Spin Facilitated Models} \label{subsec:fsm}
Spin Facilitated Models (SFMs) are a class of KCMs that exhibit dynamical heterogeneities and slowing down of relaxation timescales~\cite{rev:fsm}. 
Even though they can be implemented in a variety of lattice geometries and dimensions, from now on we will limit ourselves to $1d$. 
Let us now consider the energy function of non-interacting spins in an external field, 
\begin{equation} \label{FA}
E = K\sum_{i=1}^N n_i,
\end{equation}
introduced by Fredrickson and Andersen in Refs.~\cite{fam,fam1}, where $n_i=0,1$ is a classical spin variable. In our setup, $n_i=0 \, (1)$ corresponds to the $i$-th atom being in a \textit{bright} ({\it dark}) period.
The energy function in Eq.~(\ref{FA}) does not feature a finite temperature phase transition for a field strength $K\neq 0$. The equilibrium concentration of defects, which we define as spins in
the $n_i = 1$ state, reads 
\begin{equation} \label{eqconc}
d_{\mathrm{eq}} = \frac{1}{N}\lim_{t \rightarrow \infty} \sum_{i=1}^N n_i (t) = \frac{1}{1+e^{\beta K}},
\end{equation}
where $\beta = 1/(k_B T)$, $k_B$ is the Boltzmann's constant and $T$ is the equilibrium temperature of the system. Furthermore, for every $K>0$, $\lim_{T\rightarrow 0} d_{\mathrm{eq}} = 0$, in line with the intuition that at low $T$ the concentration of defects should be small. From now on, without loss of generality, we set $k_B =1 $ and we fix the temperature scale setting $K=1$. 

Despite the simple ground-state of Eq.~(\ref{FA}), the non-equilibrium dynamics towards it can be non-trivial. 
In general, the evolution of a classical system of spins is described by the following rate equation for the probabilities $p(\textbf{n},t)$ of being in the many-body spin state specified by the vector $\textbf{n} = (n_1, ..., n_i, ..., n_N )$ at time $t$:

\begin{equation} \label{master}
\frac{\partial p(\textbf{n},t)}{\partial t} = \sum_{\textbf{m} \neq \textbf{n}} \left \{ \Gamma(\textbf{m} \rightarrow \textbf{n}) \, p(\textbf{m},t) - \Gamma(\textbf{n} \rightarrow \textbf{m}) \, p(\textbf{n},t)  \right \}.
\end{equation}
Here, $\Gamma(\textbf{m} \rightarrow  \textbf{n})$ is the rate of the transition from  $\textbf{m}$ to $\textbf{n} \neq \textbf{m}$. The  equilibrium probabilities $\pi (\textbf{m},t)$ satisfy the detailed balance $\Gamma(\textbf{m} \rightarrow \textbf{n}) \, \pi (\textbf{m},t) = \Gamma(\textbf{n} \rightarrow  \textbf{m}) \, \pi (\textbf{n},t)
$ with respect to Eq.~(\ref{FA}). 
Under the assumption of {\it{local}} spin-flips and Glauber dynamics~\cite{glauber}, the transition rate from $(n_1,...,n_i,...,n_N)$ to $(n_1,...,1-n_i,...,n_N)$ at site $i$, $\Gamma^{n_i \rightarrow 1-n_i}$,  is equal to $1/(1 + e^{ \beta \Delta E})$, with $\Delta E =1-2n_i$ the energy difference between the final and initial spin configurations. Since in our case $\Delta E = \pm 1$, from Eq.~(\ref{eqconc}) single-particle rates can be written as
\begin{equation} \label{woutc}
\Gamma^{n_i \rightarrow 1-n_i} = (1-d_{\mathrm{eq}})n_i + d_{\mathrm{eq}} (1-n_i).
\end{equation}
Thus, the asymmetry between the ratio of the rates of creation over destruction of a defect 

\begin{equation} \label{ratior} 
\frac{\Gamma^{0 \rightarrow 1}}{\Gamma^{1 \rightarrow 0}} = \frac{d_{\mathrm{eq}}}{1-d_{\mathrm{eq}}},
\end{equation}
fixes the temperature $T$ at which the system of non-interacting spins thermalizes.  (Note that $\Gamma^{0 \rightarrow 1} \rightarrow \Gamma^{1 \rightarrow 0}$ for $T\rightarrow \infty$.) 

Up to now, all transitions between different configurations are allowed. The key idea behind SFMs is to introduce
restrictions in these transitions. The general rule can be stated as: a spin flip is allowed only if there are enough defects in the neighbourhood that can facilitate the process. 
In the 1$d$ 1-SFM, these so-called kinetic constraints require that a spin can flip provided \textit{at least} one of its two neighbors is a defect. 
This facilitation condition transforms Eq.~(\ref{woutc}) into
\begin{equation} \label{facrate}
\Gamma^{n_i \rightarrow 1-n_i} = (n_{i-1} + n_{i+1}) \, \{(1-d_{\mathrm{eq}})n_i + d_{\mathrm{eq}} (1-n_i)\}.
\end{equation}

The association between rates in Eq.~(\ref{facrate}) and the many-body rates calculated in Subsec.~\ref{many_body} is straightforward: $n_{i-1} = n_{i+1} = 0$ corresponds to $\ell=0$ for which no spin flip at site $i$ is allowed, while either $\{n_{i-1} = 1$, $n_{i+1} = 0\}$ or $\{n_{i-1} = 0$, $n_{i+1} = 1\}$ corresponds to $\ell=1$ and $i$ is facilitated to flip its spin. The only difference arises in the case of $\{ n_{i-1} = n_{i+1} = 1 \}$: while rates in the 1-SFM  are doubled with respect to the case $\{n_{i-1} = 1$, $n_{i+1} = 0\}$, for $\ell=2$, rates are suppressed as much as in the case of $\ell=0$. This will lead to different steady states which we will discuss in Sec.~\ref{sec:num_res}.

\section{Numerical results} \label{sec:num_res}
\subsection{Semi-classical regime} \label{subsec:withoutd}
\begin{figure}[h!] 
\centerline{\includegraphics[width=1.0\textwidth]{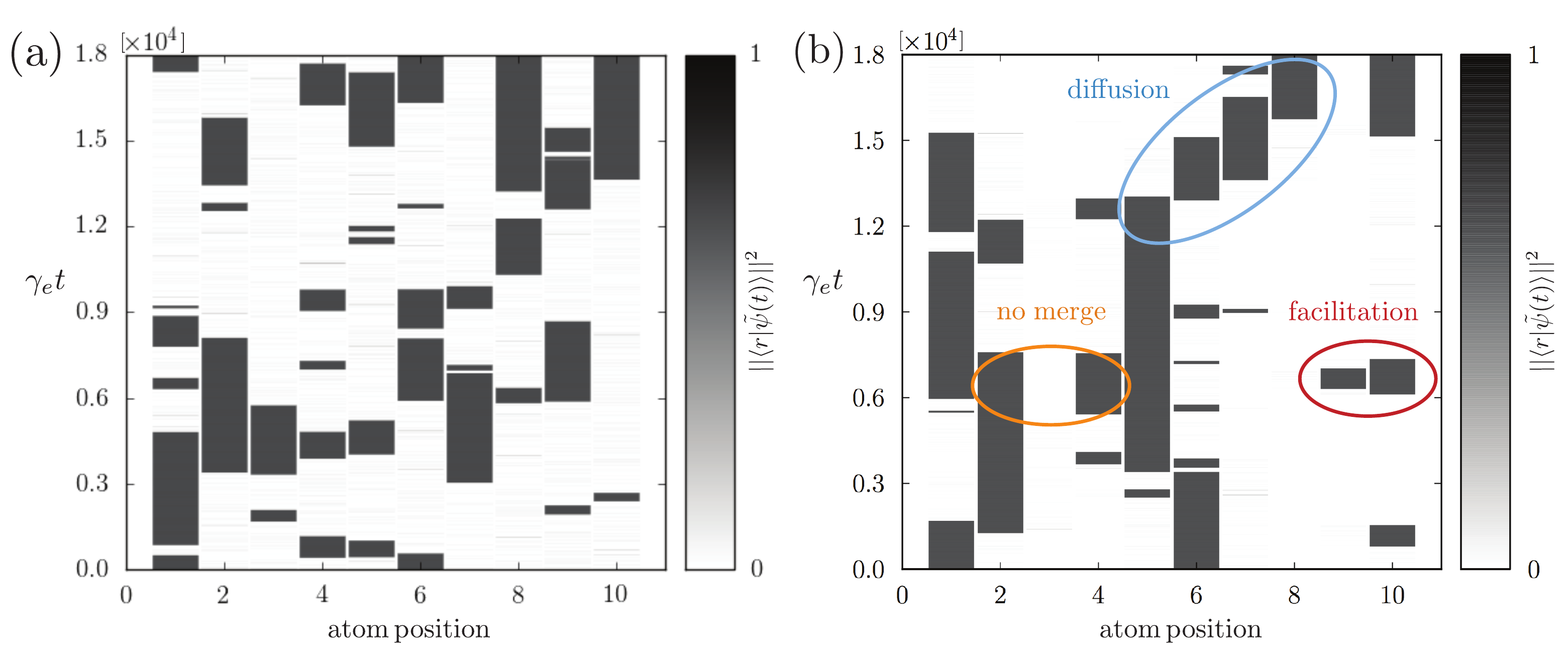}}
\caption{Comparison of a typical trajectory of Rydberg excitations in the unconstrained case $\Delta_r = V = 0$ [panel (a)] and facilitated case $\Delta_r = V = 10\gamma_e$ [panel (b)]. Other parameters are: $N = 10, \Omega_e = \gamma_e$, $\Omega_r = 0.03\gamma_e$. The initial condition is two Rydberg excitations in $i=1$ and $i=6$, while all other atoms are in $|g\rangle$. Processes of facilitation and diffusion (typical of 1-SFMs) and avoided merging (not present in 1-SFMs)  are highlighted in panel (b).}
\label{fig:traj122}
\end{figure}

In the limit of strong driving to the intermediate state $\Omega_e \sim \gamma_e$, the dynamics of the system is dominated by dissipation. As a method to study the system in the presence of both coherent driving and dissipation, we use quantum trajectory simulations (cfr.~\ref{appendix:st}). 
Fig.~\ref{fig:traj122} depicts a typical trajectory of the Rydberg population $r_{\alpha}(t)$ of several atoms [Eq.~\eref{singler}]. While for $\Delta_r = V = 0$ [panel (a)], the dynamics of Rydberg excitations is unconstrained and all atoms are independent from each other, for $\Delta_r = V \neq 0$ [panel (b)] clear spatial correlations emerge and the dynamics of Rydberg excitations resembles the dynamics of defects in the 1$d$ 1-SFM~\cite{rev:fsm,leon_berth}. This dynamics can be intuitively understood from three different species of atoms, which we label, in analogy to the language of SFMs, as:
\begin{itemize}
\item \textit{defects}, that is atoms in $|r\rangle$; 
\item \textit{facilitated} atoms, that is atoms in either $|g\rangle$ or $|e\rangle$ with one \textit{defect} as nearest-neighbor;
\item \textit{non-facilitated} atoms, that is atoms in either $|g\rangle$ or $|e\rangle$ with no \textit{defect} as nearest-neighbor.
\end{itemize}
The dynamics of defects in the $1d$ 1-SFM is known to be mainly diffusive~\cite{leon_berth}. Let us illustrate the diffusion process of {\it defects} in more detail. A typical trajectory is shown in Fig.~\ref{fig:traj122}(b) and marked by a blue ellipsis. A \textit{defect} can facilitate the excitation of a neighboring atom. In this example, the atom to the right is excited. For a short time, both are in the excited state but soon the original atom is de-excited: therefore, the \textit{defect} has effectively moved to the right by one site. 
Another  possible move is highlighted in Fig.~\ref{fig:traj122}(b) with a red ellipsis. In this second example, the left neighbor is excited but later on immediately de-excited. Here, the \textit{defect} did effectively not move. 
The diffusion process can also be understood quantitatively in some limits. In the regime $T\ll 1$ (or, equivalently, $d_{\mathrm{eq}} \ll 1$), 
 $\Gamma^{0\rightarrow 1}(0)/\Gamma^{1\rightarrow 0}(0) \sim d_{\mathrm{eq}} \sim e^{-\beta}$ and the diffusion constant is   $d_{\mathrm{eq}}/2$~\cite{rev:fsm}. In the intermediate regime we studied in this work, the diffusion constant does not have such a simple analytical expression, since $\Gamma^{0\rightarrow 1}(0)/\Gamma^{1\rightarrow 0}(0) = 0.33$.

\begin{figure}[h!] 
\centerline{\includegraphics[width=0.8\textwidth]{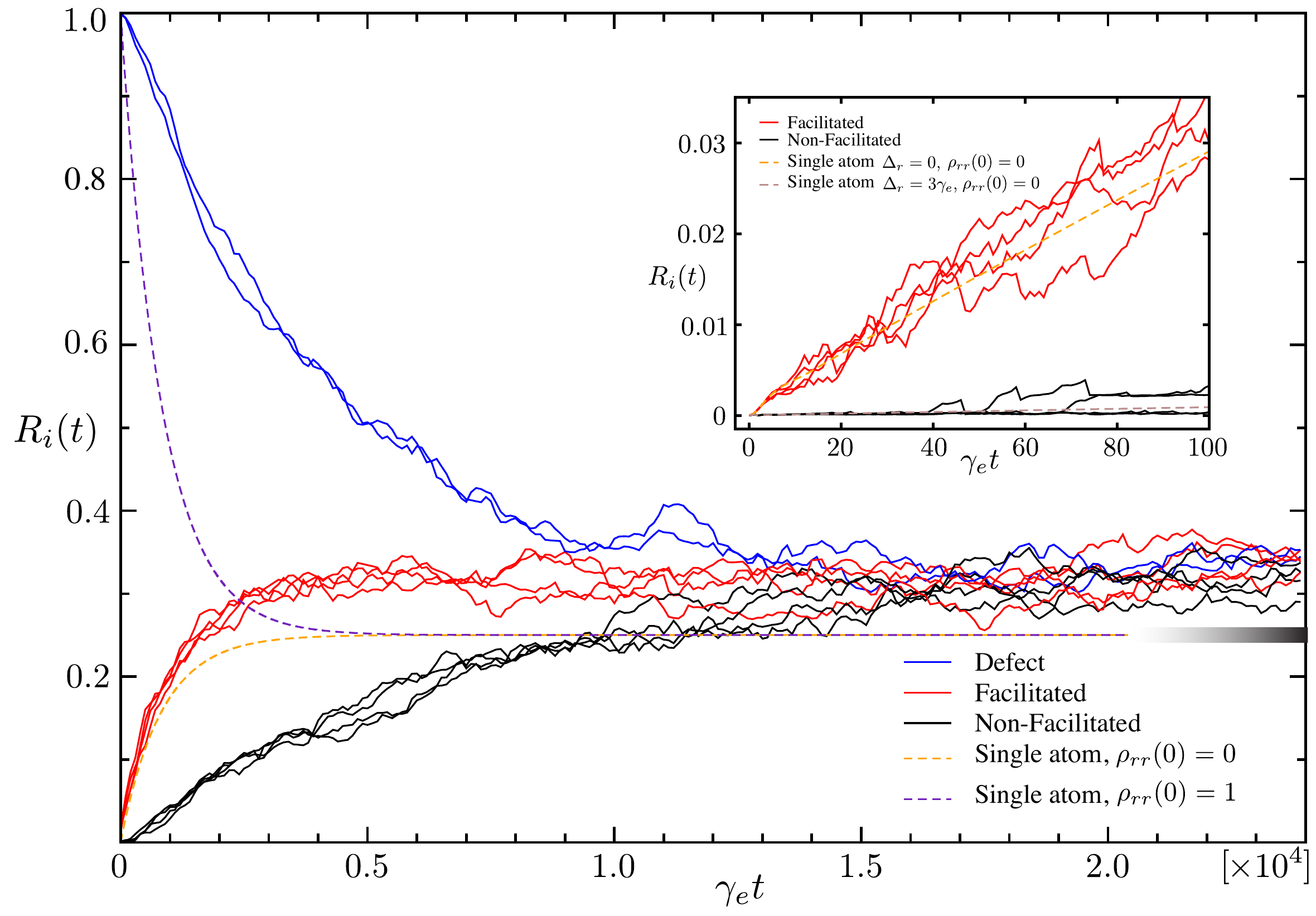}}
\caption{Evolution of the trajectory averaged Rydberg population $R_i(t)$ for $N_{\mathrm{traj}} = 500$ in a chain of $N=10$ sites. Parameters are: $\Omega_e = \gamma_e$, $\Omega_r = 0.03\gamma_e$, $\Delta_r = V = 3\gamma_e$, $\Delta^{\prime}_{ri}=0$. The initial state is: \textit{defects} (blue lines) at $i=1$ and $i=6$; \textit{facilitated} atoms (red lines) at $i=2,5,7,10$ and  \textit{non-facilitated} atoms (black lines) at $i=3,4,8,9$. Exact master equation evolution of the single atom Rydberg population $\rho_{rr}(t)$ are shown for the initial states $\rho_{rr}(0) = 0$  (orange dotted line) and $\rho_{rr}(0) = 1$ (indigo dotted line). The result $r>d_{\mathrm{eq}}$, suggests that our system lacks of a true thermalisation to the ground-state of the 1-SFM (see main text). The shaded thick black line is the value of $d_{\mathrm{eq}} = 0.25$ according to Eq.~(\ref{deq}). Inset: Resonant (orange dotted line) and off-resonant (light brown dotted line) profiles $\rho_{rr}(t)$ of a single atom well approximate the short time dynamics of \textit{facilitated} and \textit{non-facilitated} atoms, respectively.}
\label{fig:fardefect}
\end{figure}

In the following, we study the relaxation of the three different species individually. Numerical simulations of $N_{\mathrm{traj}}$ stochastically different trajectories are averaged in order to approximate the dynamics of the system described by the master equation~(\ref{ffme}). Results plotting the trajectory averaged Rydberg population $R_i(t)$ (see~\ref{appendix:st}) for each atom $i$ are depicted in Fig.~\ref{fig:fardefect}. 
\textit{Facilitated} atoms (red lines) show a dynamical behaviour analogous to the one of resonantly driven single atoms, due to the anti-blockade condition which acts as a facilitating term in the 1-SFM [cfr. Eq.~(\ref{facrate})]. In fact, the inset of Fig.~\ref{fig:fardefect} shows that, in proximity of $t=0$, the profiles of \textit{facilitated} atoms are very well approximated by the resonant ($\Delta_r = 0$) single-atom Rydberg population $\rho_{rr}(t)$ (see~\ref{appendix:st}), the associated transition rate being $\Gamma^{0 \rightarrow 1}(0)$.

Contrarily, the dynamics of \textit{defects} (blue lines)  and \textit{non-facilitated} atoms (black lines) is slowed down with respect to the one of \textit{facilitated} atoms. 
At  $t=0$, the weak transition $|g\rangle \leftrightarrow |r\rangle$ of  a \textit{defect} is  far-off-resonant and $\Gamma^{1\rightarrow 0} (\Delta_r)$ is orders of magnitude suppressed with respect to $\Gamma^{1\rightarrow 0} (0)$  [cfr. Tab.~\ref{tab:rates_off}]. A \textit{defect} can in fact relax only after one of its neighboring \textit{facilitated atoms} is excited to the Rydberg state. Similarly,  at $t=0$,
the weak transition $|g\rangle \leftrightarrow |r\rangle$ of  a \textit{non-facilitated} atom is  far-off-resonant and $\Gamma^{0\rightarrow 1} (\Delta_r)$ is orders of magnitude suppressed with respect to $\Gamma^{0\rightarrow 1} (0)$.
The inset of Fig.~\ref{fig:fardefect} shows that profiles of \textit{non-facilitated} atoms agree with the far-off resonant ($\Delta_r \neq 0$) single atom Rydberg population $\rho_{rr}(t)$: \textit{non-facilitated} atoms, in order to be excited to the Rydberg state, have to wait for one of their two neighbors to be Rydberg-excited too.
Note that for larger  $N$ we predict the existence of \textit{non-facilitated} atoms with profiles of $R_i(t)$ even slower than the ones in Fig.~\ref{fig:fardefect}. This can be regarded as a form of hierarchical relaxation~\cite{lesanov_kc}, in that the more distant a \textit{non-facilitated} atom is with respect to a \textit{defect} at $t=0$, the more slowed-down the evolution of $R_i(t)$ will be.

\begin{figure}[h!] 
\centerline{\includegraphics[width=0.8\textwidth]{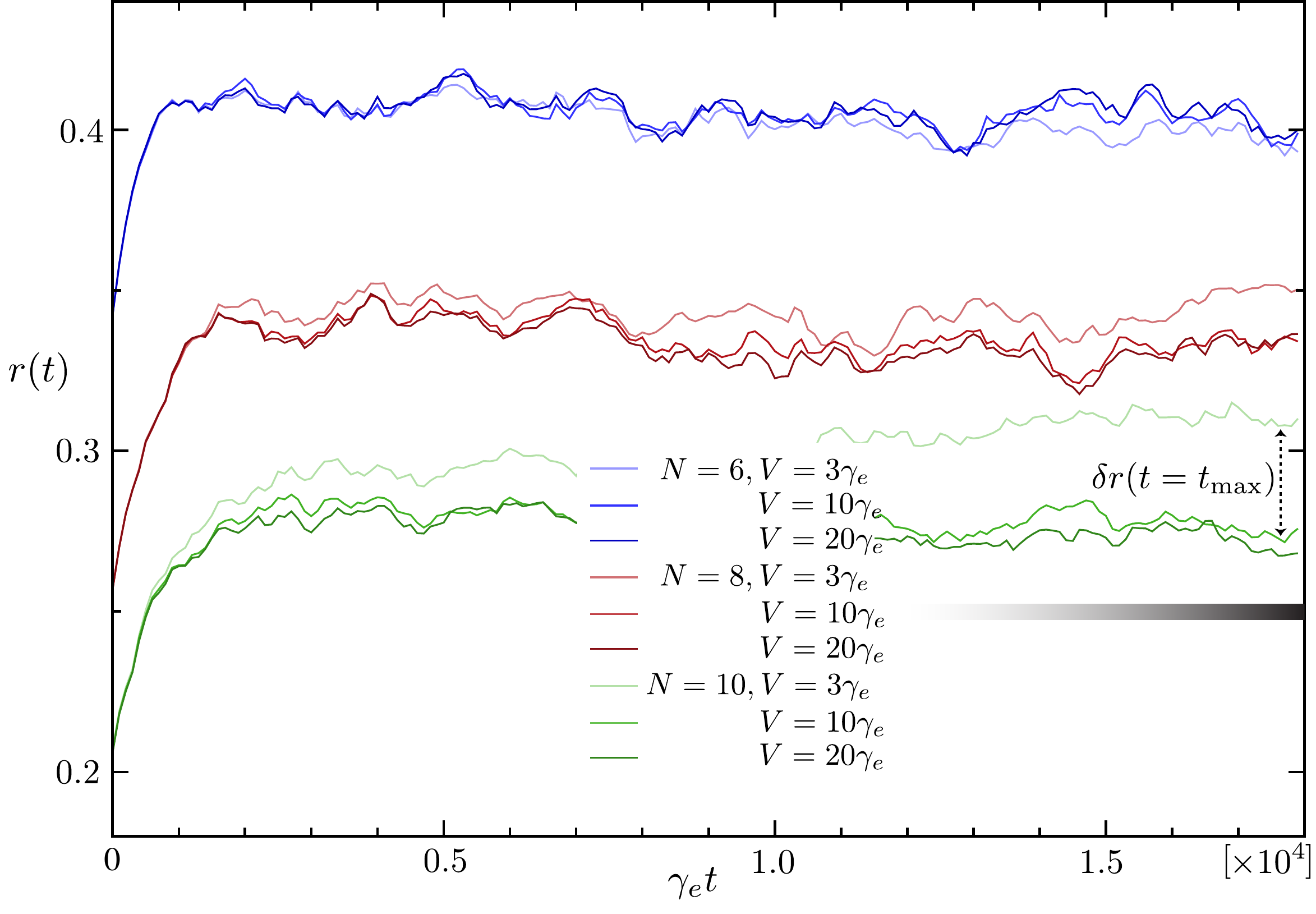}}
\caption{Evolution of the concentration of Rydberg excitations $r(t)$ for $N=(6,8,10)$, at different interaction strengths, $V=(3,10, 20) \gamma_e$. Parameters are the same as in Fig.~\ref{fig:fardefect} and the initial state is, for all plots, two \textit{defects} as far as possible from each other: assuming PBCs, for $N=6$ the initial largest distance is $3$, for $N=8$ is $4$, and for $N =10$ is $5$. The length of the dotted line agrees with $\delta r(t=t_{\mathrm{max}})$ of the main text.}
\label{fig:fsize}
\end{figure}

We now consider the concentration of Rydberg excitations  
\begin{equation} \label{cRyd}
r(t) = \frac{1}{N}\sum_{i=1}^N R_i(t).
\end{equation}
The dynamics of $r(t)$ for different $N$ and $V$, starting with the same initial state  is depicted in  Fig.~\ref{fig:fsize}. 
After an initial transient relaxation, $r(t)$ reaches a steady state constant value $r \equiv \lim_{t\rightarrow \infty} r(t)$, 
except in the case of the largest system size $N=10$ and smaller interaction  strength $V = 3\gamma_e$, where it increases in time with a small but clear trend. We determine the origin of this anomalous slow growth of $r(t)$ in the excess number of \textit{defects} created from \textit{non-facilitated} atoms within (originally suppressed) $\ell = 0$ processes, which give a negative temperature contribution in the SFM language [compare Eq.~(\ref{eqconc}) with Eq.~(\ref{ratior})]. The reason why all reverse $\ell=0$ processes of destruction of isolated Rydberg excitations are not relevant, is in the asymmetry  between off-resonant rates $\Gamma^{0 \rightarrow 1}(\Delta_r) > \Gamma^{1 \rightarrow 0}(\Delta_r)$, which increases with $|\Delta_r|$, as depicted in the inset of Fig.~\ref{fig:rand}. More quantitatively, the difference $\delta r(t)$ between the excess in the concentration of Rydberg excitations at time $t$ and the steady state value $r$ can be estimated as

\begin{equation} \label{eq:d}
\delta r(t) = \frac{N_{\mathrm{nf}} \cdot \Gamma^{0 \rightarrow 1}(\Delta_r)  \cdot  t}{N}, 
\end{equation}
where $N_{\mathrm{nf}}$ is the number of \textit{non-facilitated} atoms at $t=0$, which we approximate constant in time and equal to its value in the initial state.
In our case, at the final time of the trajectory $t_{\mathrm{max}}$, $\delta r(t_{\mathrm{max}}) \sim 0.06$, which well agrees with numerical results plotted in Fig.~\ref{fig:fsize}, within statistical errors due to finite $N_{\mathrm{traj}}$ simulated.
Contrarily, the difference $-\delta r^{\prime}(t)$ between the lack in the concentration of Rydberg excitations at time $t$ and the steady state value $r$ is

\begin{equation} \label{eq:d1}
-\delta r^{\prime}(t) = \frac{N_{\mathrm{d}} \cdot \Gamma^{1 \rightarrow 0}(\Delta_r)  \cdot  t}{N}, 
\end{equation}
where $N_{\mathrm{d}}$ is the number of  isolated \textit{defects} at $t=0$, which we also assume constant in time. Within our parameters, $\delta r^{\prime}( t_{\mathrm{max}}) \sim 0.003$, which is negligible with respect to both finite $N_{\mathrm{traj}}$ errors and $\delta r(t)$.
From Eqs.~(\ref{eq:d}), and~(\ref{eq:d1}) it is clear that, for $\Delta_r$ sufficiently large, $l=0$ off-resonant processes are inhibited due to the suppression of $\Gamma^{0 \rightarrow 1}(\Delta_r)$ and $\Gamma^{1 \rightarrow 0}(\Delta_r)$ and consequently $r(t)$ levels-off, reaching a steady state (see $\Delta_r \gtrsim 10 \gamma_e$).
All far-off-resonant $\ell=2$ processes in Fig.~\ref{fig:fsize} of merging and splitting of Rydberg excitations can be limited by a 'clever' choice of the initial concentration $r(0)$ and configuration of Rydberg excitations. This is the case for, e.g., the initial states of one \textit{defect} or two far \textit{defects} in Fig.~\ref{fig:fsize}, in which the initial number $N_{\mathrm{b}}$ of \textit{blocked} atoms is zero.  \textit{Blocked} atoms are atoms with both left and right nearest-neighbour in the Rydberg state: the choice of such initial state ensures that, most probably, $N_{\mathrm{b}} = 0$ also at later times and we can thus limit the conditions that lead to the occurrence of merging processes.

\begin{figure}[h!] 
\centerline{\includegraphics[width=1\textwidth]{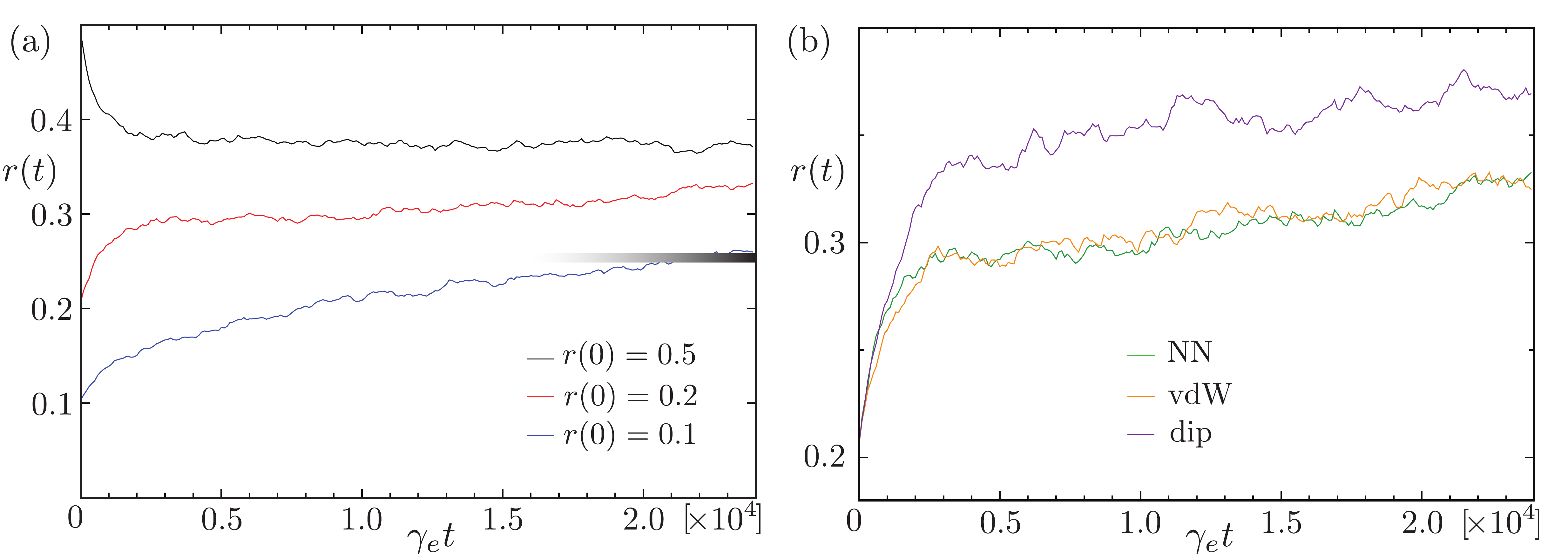}}
\caption{Panel (a): Evolution of the concentration of Rydberg excitations $r(t)$ for different initial states $r(0)$. Parameters are the same as in Fig.~\ref{fig:fardefect}. The initial configurations, assuming PBCs, are: $r(0) = 0.5$ (black line), with \textit{defects} at $i=1,2,5,7,8$; $r(0) = 0.2$ (red line), with \textit{defects} at $i=1$ and $i=5$; $r(0) = 0.1$ (blue line), one \textit{defect} (due to PBCs, its position is irrelevant). 
For all initial states, the relaxation of $r(t)$ does not converge to $d_{\mathrm{eq}}$. The best agreement with the predicted $d_{\mathrm{eq}}$ is obtained for the lowest initial concentration of Rydberg excitations, where processes of merging (see Fig.~\ref{fig:traj122}) are less probable to occur.
Panel (b): Evolution of the concentration of Rydberg excitations for the initial state $r(0) = 0.2$ of panel (a) with nearest-neighbour (NN), van der-Waals (vdW) and dipolar (dip) interaction potentials. For vdW and dip potentials, the range of pairwise interactions have been truncated to third-nearest neighbour, which we verified to be a good approximation.}
\label{fig:cinconf}
\end{figure}

Plateaus in $r(t)$ at long times in Fig.~\ref{fig:fsize} show that a steady state is always characterized by the general result $r>d_{\mathrm{eq}}$, where

\begin{equation} \label{deq}
d_{\mathrm{eq}} = \frac{\gamma_e^2}{2(\gamma_e^2+\Omega_e^2)},
\end{equation} 
and not $r=d_{\mathrm{eq}}$, as expected from the theory of 1-SFM. 
(It can be verified that $d_{\mathrm{eq}}$, in the limit $\Omega_r \ll \Omega_e, \gamma_e$, is well approximated by the single-atom steady state Rydberg population $\rho_{rr}(t \rightarrow \infty)$ calculated in~\ref{appendix:ss}. This confirms that the equilibrium ground-state of the 1-SFM is trivial, despite its spatially correlated and slowed down transient dynamics.)
Our system thus does not thermalize in the long-time limit, but it converges to a different steady state. 
Nevertheless, the trend in Fig.~\ref{fig:fsize} suggests that for larger $N$, $r$ decreases approaching $d_{\mathrm{eq}}$. 

Furthermore let us note that, due to finite size effects, $r$ crucially depends on the initial configuration of Rydberg excitations, as depicted in Fig.~\ref{fig:cinconf}(a): while the best approximation of the $1d$ 1-SFM result is given, as expected, by the lowest initial concentration of \textit{defects} (blue line), for higher initial concentrations (black line), avoided merging processes become relevant and other corrections to $r(t)$, similar to those in Eqs.~(\ref{eq:d}) and~(\ref{eq:d1}), come into play. 
Contrarily, in the case in which both $\ell = 2$ merging and splitting processes are present and in the limit of large $N$, our system would thermalize for every initial state in the long-time limit.

We finally mention that, ideally, one could think of simulating the full dynamics of the $1d$ 1-SFM introducing a third laser which weakly  couples $|g\rangle \leftrightarrow |r\rangle$ with a small Rabi frequency $\Omega_r$ and detuning $2 \Delta_r$ from $|r\rangle$. In this configuration,  
processes of merging of Rydberg excitations would be allowed also for the finite $N$ we simulated numerically.

\subsection{Quantum Regime} \label{subsec:sa_re}
In the limit of vanishing driving of the $|g\rangle \leftrightarrow |e\rangle$ transition,  $\Omega_e \ll \gamma_e, \Omega_r$,  the dynamics, up to times $ t \ll \gamma_e/ \Omega_e^2$, is dominated by  $|g\rangle \leftrightarrow |r\rangle$ coherences.
Starting from the fully coherent limit $\Omega_e = 0$, one can study, increasing $\Omega_e$, the crossover to the classical rate equation limit of Subsec.~\ref{subsec:withoutd}, i.e. $\Omega_e \sim \gamma_e$ with $\Omega_r \ll \Omega_e^2/\gamma_e$.

In the case of $\Omega_e=0$, the system Hamiltonian is that of $N$ interacting two-level atoms
\begin{equation}  \label{coheham}
H_{\mathrm{tl}} = \sum_{i=1}^N \Biggl \{ \frac{\Omega_r}{2} \bigg(|g \rangle \langle r |_i + \mathrm{H.c.}\bigg) - \Delta_r |r \rangle \langle r |_i \Biggr \} + V \sum_{i=1}^{N-1} \biggl \{ |r \rangle \langle r |_i \otimes |r \rangle \langle r |_{i+1}  \biggr \}.
\end{equation} 
The aim here is to study the non-equilibrium dynamics of the average concentration of Rydberg excitations $r(t)$.  
The Hamiltonian in Eq.~(\ref{coheham}), assuming $V = \Delta_r$ and PBCs, can be mapped exactly to the so-called transverse field Ising model~\cite{sachdev}. We note that the time evolution of the magnetization, which in our formalism can be easily related  to $r(t)$, following a quantum quench within the ordered (ferromagnetic) phase and across the phase transition to the disordered (paramagnetic) phase, has been extensively studied in Refs.~\cite{calabrese1} and~\cite{calabrese2}. There, the authors provide explicit analytical results in the thermodynamic limit $N \rightarrow \infty$, corroborated by numerical computations showing that finite-size effects are negligible.

\begin{figure}[h!] 
\centerline{\includegraphics[width=0.8\textwidth]{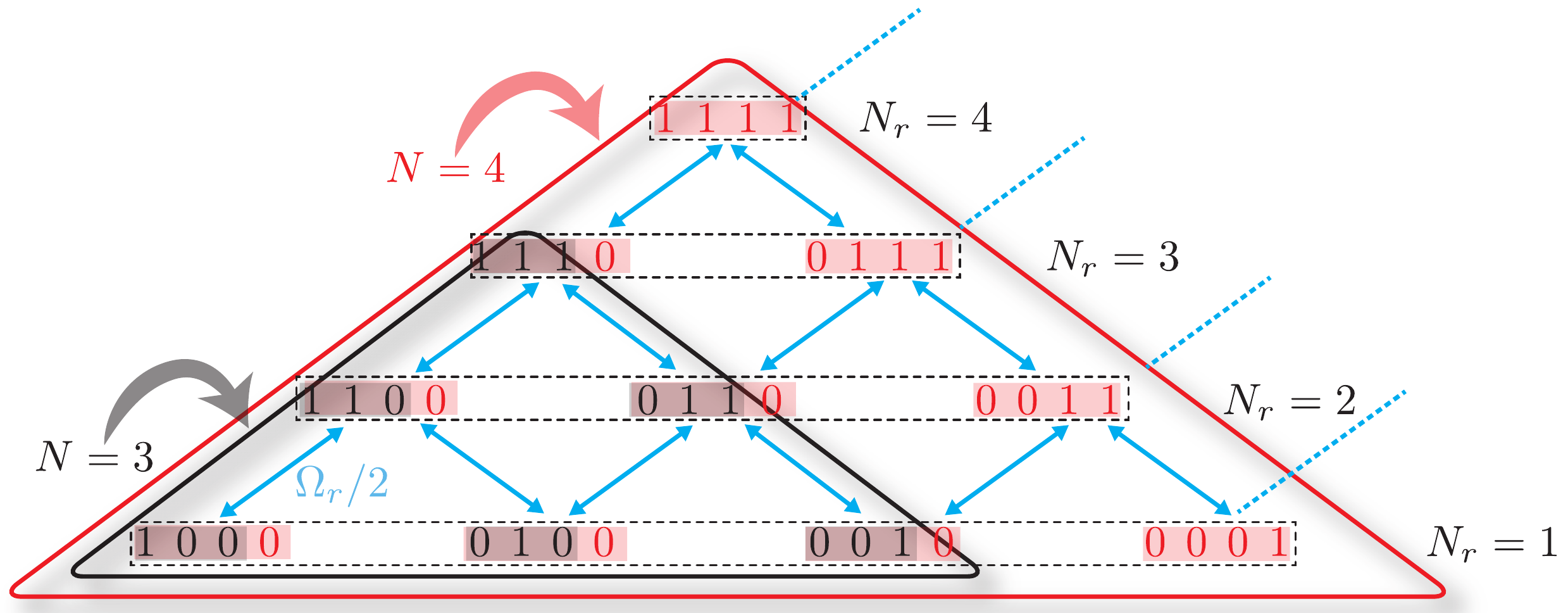}}
\caption{Scheme of the effective dynamics  governed by the Hamiltonian in Eq.~(\ref{coheham}) with open boundary conditions (OBCs) for $t \ll \Delta_r/\Omega_r^2$, given an initial state with energy $-V$ (see main text). The binary values $0$ and $1$ represent here an atom in $|g\rangle$ and $|r\rangle$, respectively. All allowed system configurations are coupled with the single atom Rabi frequency $\Omega_r/2$ (blue arrows). 
The black (red) triangle refers to the case with $N=3$ ($N=4$) atoms. In each row, the total number of Rydberg excitations $N_r$ is constant, independently of $N$.
Increasing the number of atoms from $N=3$ to $N=4$, results in adding a diagonal group of $N=4$ new states (highlighted in shaded red) and adding a $0$ in the rightmost site of all $N=3$ states (highlighted in shaded black). This scheme can be generalized to an arbitrary $N$ iteratively.}
\label{fig:tree}
\end{figure}

In the following, we show that for $V=\Delta_r$ {\it and} in the limit $\Omega_r \ll \Delta_r$, one can approximate the dynamics of excitations of the transverse field Ising model [or of Eq.~(\ref{coheham})], whose Hilbert space scales exponentially ($2^N$), with the dynamics of excitations in a simplified model with only polynomial scaling ($N^2$).  The Hilbert space of such a simplified model (Fig.~\ref{fig:tree}) is just a subspace of the transverse field Ising model total Hilbert space, and is univocally determined by the initial state of the dynamics.

Let us derive this simplified model from an illustrative example. Considering $N =3$ atoms, $H_{\mathrm{tl}}$ can be represented, in the basis $\{|ggg\rangle, |rgr\rangle, |ggr\rangle, |grg\rangle, |grr\rangle, |rgg\rangle, |rrg\rangle, |rrr\rangle \}$, by the matrix 

\begin{equation} \label{redmatrix}
\mathcal{H}^{N=3}_{\mathrm{tl}}  =
\left(\begin{array}{cc;{2pt/2pt}cccccc}                
                        0 & 0 & \frac{\Omega_r}{2} &  \frac{\Omega_r}{2} & 0 &   \frac{\Omega_r}{2}  & 0  & 0      \\
                        0 & -2V & \frac{\Omega_r}{2} & 0   & 0 &   \frac{\Omega_r}{2} & 0  & \frac{\Omega_r}{2} \\ \hdashline[2pt/2pt]
                        \frac{\Omega_r}{2} &  \frac{\Omega_r}{2} & -V &  0 & \frac{\Omega_r}{2}  & 0 & 0  & 0 \\ 
                        \frac{\Omega_r}{2} & 0 & 0 & -V   & \frac{\Omega_r}{2}  &  0 & \frac{\Omega_r}{2}  & 0 \\
                       0 & 0 &  \frac{\Omega_r}{2}&  \frac{\Omega_r}{2}   &   -V &  0 & 0  & \frac{\Omega_r}{2} \\
                       \frac{\Omega_r}{2} & \frac{\Omega_r}{2} & 0 & 0   & 0   &  -V  & \frac{\Omega_r}{2}  & 0 \\
		      0 & 0 &  0 & \frac{\Omega_r}{2}   & 0  &  \frac{\Omega_r}{2} &  -V   & \frac{\Omega_r}{2} \\
		      0 & \frac{\Omega_r}{2} & 0 & 0   & \frac{\Omega_r}{2}  &  0 & \frac{\Omega_r}{2}   & -V  

\end{array}\right).
\end{equation}
If the initial state has energy $-V$, the dynamics of the system is limited, for times $t \ll \Delta_r/\Omega_r^2$, to the subspace $|rrr\rangle, |ggr\rangle, |grg\rangle, |grr\rangle, $ $|rgg\rangle, |rrg\rangle$ of the full Hilbert space. The matrix $\tilde{\mathcal{H}}^{N=3}_{\mathrm{tl}}$,  enclosed within the dotted lines in the down-right corner of $\mathcal{H}^{N=3}_{\mathrm{tl}}$ in  Eq.~(\ref{redmatrix}), governs the dynamics of the system in the reduced subspace. 

\begin{figure}[h!] 
\centerline{\includegraphics[width=0.8\textwidth]{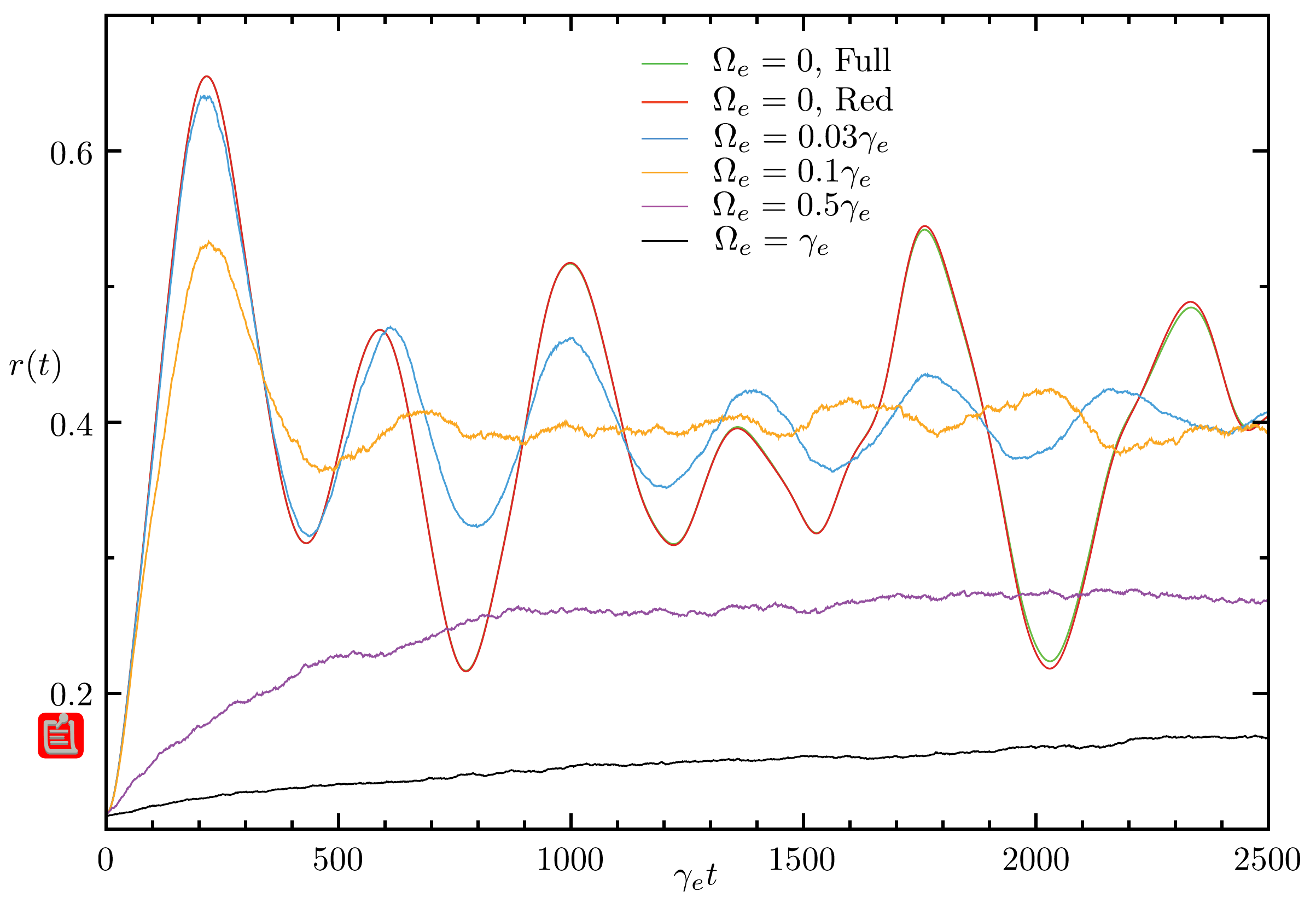}}
\caption{Evolution of $r(t)$ for different $\Omega_e$, $N = 9$, and OBCs. Hamiltonian parameters are: $\Omega_r = 0.03 \gamma_e$, $V = \Delta_r = 3\gamma_e$. The initial state is  one \textit{defect} in $i=5$, all other atoms being in $|g\rangle$. The green curve plots results of simulations evolving the full Hamiltonian $\mathcal{H}^{N=9}_{tl}$ for $\Omega_e = 0$. The red curve reproduces with high accuracy the exact dynamics evolving the reduced matrix $\tilde{\mathcal{H}}^{N=9}_{tl}$ for $\Omega_e = 0$ (see discussion in the main text). For all $\Omega_e \neq 0$, $N_{\mathrm{traj}} = 500$  trajectories have been simulated.}
\label{fig:comp}
\end{figure}

In our example, the subspace is characterized by the presence of one cluster of Rydberg excitations with varying size. Here, a cluster is a group of contiguous Rydberg excitations without atoms in $|g\rangle$ in between them. The total number of Rydberg excitations $N_r$ in the cluster changes during the dynamics (see Fig.~\ref{fig:tree}), but the number of clusters does not. At this point we remark that, at times $t \gtrsim \Delta_r/\Omega_r^2$, second (and higher) order perturbative processes in the small parameter $\Omega_r/\Delta_r $ will drive the dynamics outside the subspace. 
This discussion can be extended also for a general $N$ and for every initial condition. The number of clusters will be a constant of motion and it will always be determined by the initial state. 
For example, in Fig.~\ref{fig:comp}, we compare the evolution  of $r(t)$ governed by $\mathcal{H}^{N}_{\mathrm{tl}}$ (green line)  and $\tilde{\mathcal{H}}^{N}_{\mathrm{tl}}$ (red line) for $N=9$. 
We restrict the dynamics of the system in the subspace with energy $-V$ starting with one Rydberg excitation in the center of the open chain and all the remaining  atoms in $|g\rangle$. Numerical simulations of the full ($\mathcal{H}^{N=9}_{\mathrm{tl}}$) and effective ($\tilde{\mathcal{H}}^{N=9}_{\mathrm{tl}}$) dynamics are in excellent agreement for $t \ll \Delta_r/\Omega_r^2$. In this time window, the peaks of oscillations of $r(t)$ have a periodicity $2\pi/(\Omega_r/2$).

Remarkably, the dimension of $\tilde{\mathcal{H}}^{N}_{\mathrm{tl}}$ scales as $N(N+1)/2$ contrarily to the exponential growth $2^N$ of  $\mathcal{H}^N_{\mathrm{tl}}$. Such quadratic scaling for large $N$ is  understood from the the fact that the effective dynamics of our system, in that limit, can be interpreted as hopping processes of a single particle in a $2d$ square lattice, which is the well-known $2d$ tight-binding model. We will comment on this in a few lines.

\begin{figure}[h!] 
\centerline{\includegraphics[width=0.8\textwidth]{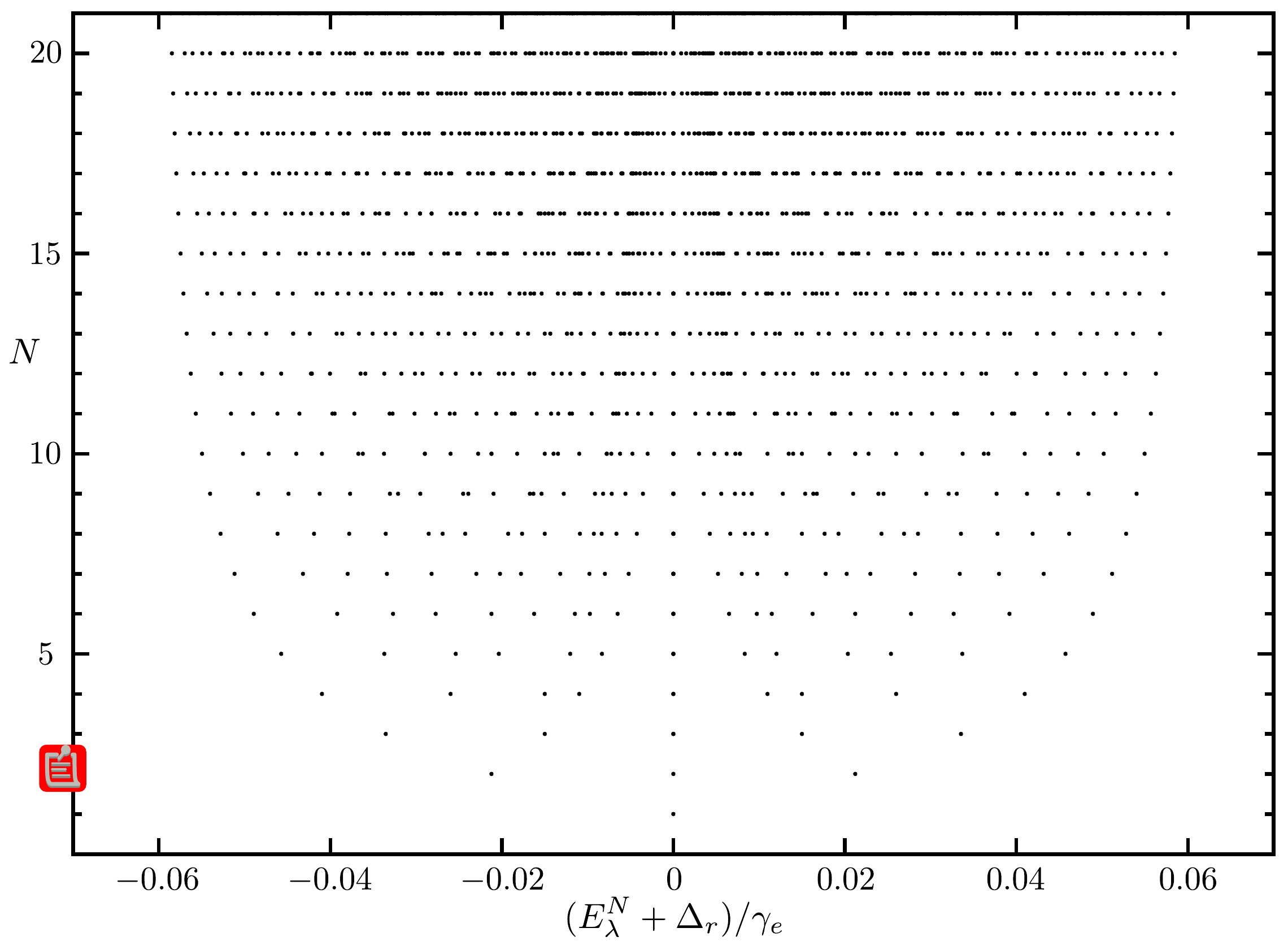}}
\caption{Eigenvalues spectrum for $\Omega_r = 0.03\gamma_e$ and $\Delta_r = 0$ at different $N$.  In the thermodynamic limit $N\rightarrow \infty$, the spectrum is bounded by $\pm 2\Omega_r$, according  to the two-dimensional tight binding model dispersion relation [see Eq.~(\ref{disp}) and the related discussion in the main text].}
\label{fig:eig}
\end{figure}

It is worthwhile first to study properties of $\tilde{\mathcal{H}}^{N}_{\mathrm{tl}}$ more in detail. Setting $\Delta_r=V=0$ without loss of generality (in the reduced subspace of $\tilde{\mathcal{H}}^{N}_{\mathrm{tl}}$, $V$ is just a constant energy off-set) and considering again the example $N=3$, the eigenvalues $E^{N=3}_{\lambda}$ of $\tilde{\mathcal{H}}^{N=3}_{\mathrm{tl}}$, written in ascending order, are 

\begin{equation} \label{eq:eig}
E^{N=3}_{\lambda} = \Bigg (-\frac{\sqrt{5} \Omega_r}{2},  -\frac{\Omega_r}{2}, 0, 0 ,   \frac{\Omega_r}{2},    \frac{\sqrt{5} \Omega_r}{2} \Bigg ).
\end{equation}
In Fig.~\ref{fig:eig}, we plot the spectrum of eigenvalues $E^{N}_{\lambda}$ for different $N$. Two features emerge:

$(i)$ Independently of $N$, there is always at least one \textit{dark} state, that is an eigenstate of $\tilde{\mathcal{H}}^{N}_{\mathrm{tl}}$ with $E^{N}_{\lambda} = 0$. We verified that the total number of degenerate \textit{dark} states increases linearly in $N$. More specifically,  for $N\geqslant 3$ [$N\geqslant 2$] odd [even], it scales with $(N+1)/2$ [$N/2$]. 
In the case of $N=3$, the two (non-normalized)  \textit{dark} states can be expressed as
\begin{equation} \label{dark}
|D^{N=3}\rangle =  |rrr\rangle - |grg\rangle \hspace{0.5cm} \mbox{and} \hspace{0.5cm}  |E^{N=3}\rangle = |rgg\rangle - |grg\rangle + |ggr\rangle .
\end{equation}
In particular, $|E^{N}\rangle$ exists for every $N$, and it is obtained from the linear combination of all states consisting of only one Rydberg excitation with alternating plus and minus signs, in analogy to $|E^{N=3}\rangle$.
Due to the linear scaling in $N$ of the degeneracy of \textit{dark} states, and since bulk properties of the system scale quadratically, we can think of  \textit{dark} states also as \textit{edge} states in the Hilbert space of $\tilde{\mathcal{H}}^{N}_{\mathrm{tl}}$: for example, $|E^{N=3}\rangle$ is localized at the lower boundary of the scheme depicted in Fig.~\ref{fig:tree}.

$(ii)$ The eigenvalue energy spectrum, shown in Fig.~\ref{fig:eig} for different $N$, is bounded by $-2\Omega_r$ and $2\Omega_r$ for large $N$.  
This is a consequence of the fact that, as already mentioned, our model, in the limit $N\rightarrow \infty$, can be mapped exactly to the $2d$ tight binding model. Its dispersion relation looks like

\begin{equation} \label{disp}
\varepsilon(\vec{\textbf{q}}) = -2J \{ \cos{(q_1 s)} + \cos{(q_2 s)} \},
\end{equation}
where $\varepsilon(\vec{\textbf{q}})$ is the energy of the particle at quasi-momentum $\vec{\textbf{q}} = q_1 \vec{\textbf{b}}_1 + q_2 \vec{\textbf{b}}_2$ and $J$ is the hopping amplitude between adjacent lattice sites with spacing $s$. $\vec{\textbf{b}}_{1,2} = (2\pi/s^2)  \vec{\textbf{a}}_{1,2}$ are the vectors which span the reciprocal lattice generated by $\vec{\textbf{a}}_{1} = s(1,0)$ and $\vec{\textbf{a}}_{2} = s(0,1)$  [$(1,0)$ and $(0,1)$ is the standard basis in the Euclidean plane]. Furthermore, due to PBCs, $q_{1,2}s = 2\pi m_{1,2}/N$ and $m_{1,2} \in [-N/2, N/2]$: from Eq.~(\ref{disp}), $\varepsilon(\vec{\textbf{q}})$ has thus a maximum width of $8J$. In our case $J\equiv\Omega_r/2 $, and the spectrum width of $\varepsilon(\vec{\textbf{q}})$, $4\Omega_r$, coincides  with the one observed in Fig.~\ref{fig:eig} for large $N$.  
We stress that this mapping, which simplifies a $d-$dimensional many-body problem with nearest-neighbour interactions to a $(d+1)-$dimensional single particle problem, is exact only in the limit $N \rightarrow \infty$, where boundary effects in our model are absent, i.e. PBCs, assumed in Eq.~(\ref{disp}), hold.

Let us finally comment on the transition from the fully coherent limit to the classical limit switching on and increasing $\Omega_e$, while keeping fixed all other parameters. 
Even for small $\Omega_e = (0.03,0.1)\gamma_e$ (blue and orange lines, respectively, in Fig.~\ref{fig:comp}), the system dynamics deviates from the coherent case, since  $\Omega_e \gtrsim \Omega_r $, and oscillations of $r(t)$ are damped by dissipative processes due to the coupling to the short-lived intermediate level $|e\rangle$.  For $t \gg \gamma_e/\Omega_e^2$, $r(t)$ is driven to its steady state value $r$, which is well approximated by $\rho_{rr}(t \rightarrow \infty)$ [see~\ref{appendix:ss}].
Upon further increase of $\Omega_e$,  the rate equation limit is recovered, in which $r(t)$ increases slowly with time, oscillations are completely suppressed and, as already discussed in Subsec.~\ref{subsec:withoutd}, $r$ does not coincide with $\rho_{rr}(t \rightarrow \infty)$.

\subsection{External disorder} \label{subsec:extdis}

The external disorder, arising from static spatially-random detunings from $|r\rangle$, $\Delta^{\prime}_{ri} \neq 0$ in Eq.~(\ref{newham}), considerably changes the dynamics of the relaxation. In the quantum regime, they are responsible of the many-body localization transition~\cite{anderson,altsh,rev:mbl,huse}. Here, we focus on the effects of random detunings on the relaxation of our system
in the rate equation limit of the model. We randomly choose the values of $\Delta^{\prime}_{ri}$
 within the interval $[-A,A]$ according to a uniform distribution. Moreover, we assume $0<A \ll \Delta_r$, in order to avoid  accidental balancing of the homogeneous and random detunings, $\Delta_{ri}^{\prime} \sim \Delta_r$.
Quantum trajectories are simulated according to the following procedure: (i) construct $N_{\mathrm{rnd}}$ different sets of random detunings $( \Delta_{r1}^{\prime},..., \Delta_{ri}^{\prime}, ...,  \Delta_{rN}^{\prime} )$ to be added to the homogeneous detuning $\Delta_r$; (ii) for each random field configuration  $( \Delta_{r1}^{\prime} + \Delta_r,..., \Delta_{ri}^{\prime}+ \Delta_r, ...,  \Delta_{rN}^{\prime} + \Delta_r )$, simulate $N_{\mathrm{traj}}$ stochastically independent quantum trajectories; (iii)  average the desired observable over the total number of trajectories $N_{\mathrm{traj}}  N_{\mathrm{rnd}}$. Note that, for sufficiently large $N_{\mathrm{rnd}}$, no extra energy is added to the system, because  $[-A,A]$ is centered around $0$.

\begin{figure}[h!] 
\centerline{\includegraphics[width=0.8\textwidth]{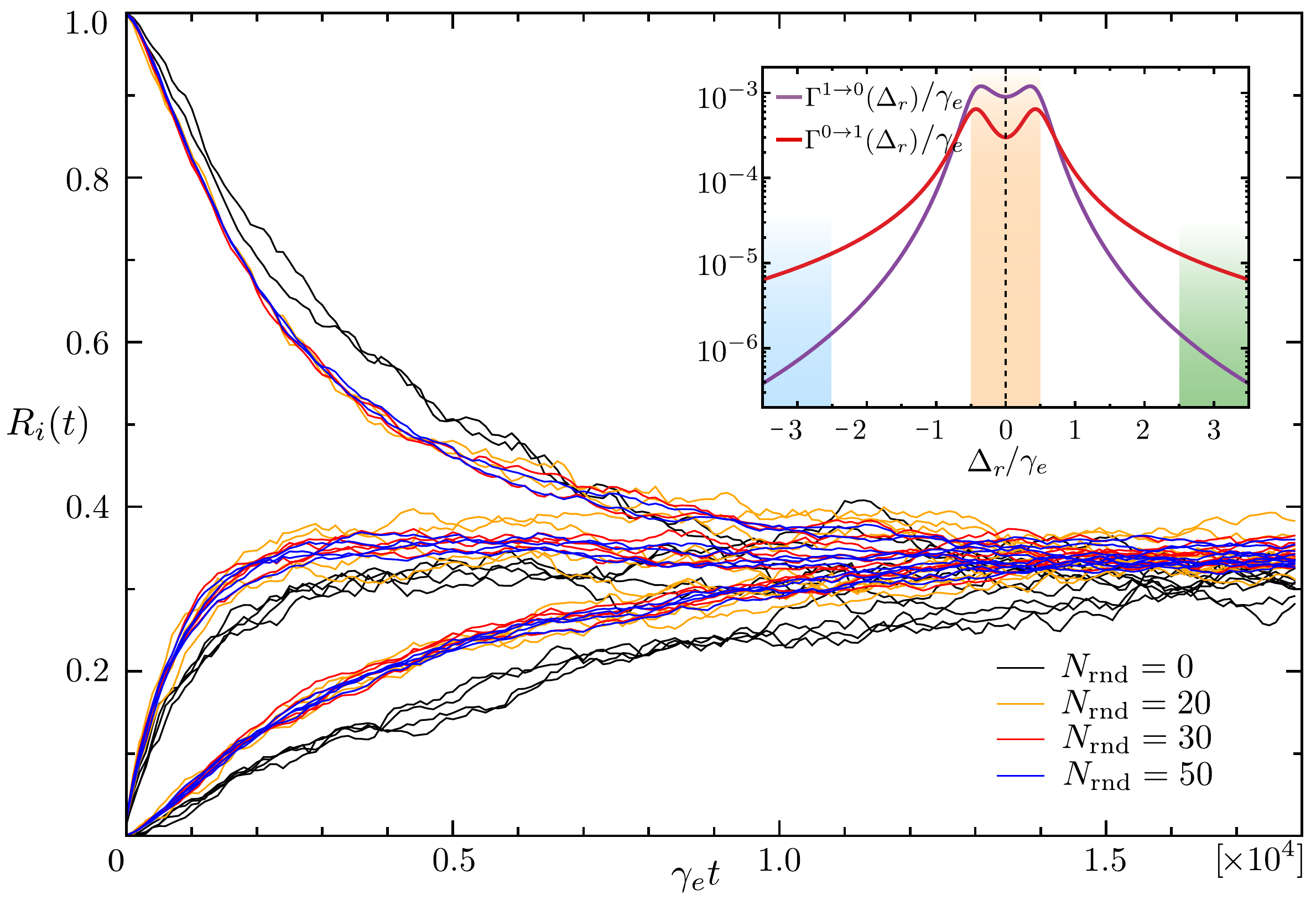}}
\caption{Evolution of the trajectory averaged Rydberg population $R_i(t)$ for different finite random detunings from the Rydberg state, $\Delta^{\prime}_{ri} \neq 0$.  The initial configuration and parameters are the same as in Fig.~\ref{fig:fardefect} and $A = 0.5\gamma_e$ (see main text). Different colors refer to different numbers of random detuning configurations $N_{\mathrm{rnd}}$, for a fixed number of trajectories simulated in each random configuration, $N_{\mathrm{traj}} = 100$.
Inset: Semi-log plot of single particle rates $\Gamma^{n_i \rightarrow 1-n_i}(\Delta_r)$ [with $n_i=0,1$] from Eqs.~(\ref{sprate}) and~(\ref{sprateup}) as a function of $\Delta_r$. The shaded areas delimit the intervals in which the total detuning $\Delta_r^{\star} \pm |\Delta^{\prime}_{ri}|$ ranges for \textit{facilitated} atoms (beige), \textit{defects/non-facilitated} atoms (green) and \textit{blocked} atoms (blue).}
\label{fig:rand}
\end{figure}

Fig.~\ref{fig:rand} shows that increasing $N_{\mathrm{rnd}}$ (for a sufficiently large $N_{\mathrm{traj}}$), all atoms relax faster toward the steady state with respect to the case of zero random detuning.  This tendency can be understood looking at Tab.~\ref{tab:rand}, which summarizes the effect of a finite random detuning on each species.

\begin{table}[h!]
\begin{tabular}{|| c | c || c | c || c | c ||}
  \hline
   \multicolumn{2}{||c||}{{\it Facilitated}: $\Delta^{\star}_r = 0$}  & \multicolumn{2}{|c||}{{\it Defect/Non-Fac.}: $\Delta^{\star}_r = \Delta_r$}  &   \multicolumn{2}{|c||}{{\it Blocked}: $\Delta^{\star}_r = -\Delta_r$} \\ \hline
   $0+|\Delta^{\prime}_{ri}|$ &$0-|\Delta^{\prime}_{ri}|$ & $\Delta_r+|\Delta^{\prime}_{ri}|$ & $\Delta_r-|\Delta^{\prime}_{ri}|$   & $-\Delta_r+|\Delta^{\prime}_{ri}|$ & $-\Delta_r -| \Delta^{\prime}_{ri}|$\\ \hline

   Speed up &Speed up & Slow down & Speed up  & Speed up &Slow down\\ \hline
\end{tabular}
\caption{Effect of  positive (first inner-columns) and negative (second inner-columns) random detunings $\Delta_{ri}^{\prime} \in [-A,A]$ (with $A \ll \Delta_r$) on \textit{facilitated} atoms, \textit{defects}, \textit{non-facilitated} atoms and \textit{blocked} atoms. The attribute 'Slow down' ('Speed up') implies that the effective rates $\Gamma^{n_i \rightarrow 1-n_i} (\Delta_r^{\star} \pm |\Delta^{\prime}_{r}|)$ [with $n_i=0,1$] are reduced (increased) with respect to zero random detuning rates, $\Gamma^{n_i \rightarrow 1-n_i} ( \Delta^{\star}_r )$.} 
\label{tab:rand}
\end{table}

As depicted in the inset of Fig~\ref{fig:rand}, for $A \ll \Delta_r$, rates with a finite random detuning $\Gamma^{n_i \rightarrow 1-n_i}(\Delta^{\prime}_{ri})$ of \textit{facilitated} atoms are larger than the corresponding rates with no random detuning $\Gamma^{n_i \rightarrow 1-n_i}(0)$ (see the two local maxima of both rates within the beige area).  As a consequence, at each time, the relaxation of \textit{facilitated} atoms tends to be sped up by any finite random detuning, independently from its sign. Contrarily, the relaxation of \textit{defects/non-facilitated} atoms  and \textit{blocked} atoms is either slowed down or sped up, depending on the sign and the magnitude of $\Delta^{\prime}_{ri}$ (see shaded green and blue areas, respectively). As a general rule, \textit{blocked} atoms and \textit{defects/non-facilitated} atoms are affected in the opposite way by the same random detuning: e.g., for $\Delta^{\prime}_{ri}>0$ \textit{defects/non-facilitated} atoms are slowed down, whereas \textit{blocked} atoms are sped up.

The result of an averaged (over trajectories and random field realizations) faster relaxation to the steady state can be understood more deeply as a consequence of the breaking of integrability of the transverse field Ising model the Hamiltonian in Eq.~(\ref{coheham}) can be mapped to.

We would like to point out that, in the different limit of large random detunings, i.e. $A \sim \Delta_r$, the dynamics is contrarily slowed down, in that \textit{facilitated} atoms have much higher probabilities of turning \textit{non-facilitated}.

\section{Experimental considerations} \label{sec:limit}

In the following, we discuss and validate the various assumptions made in Secs.~\ref{sec:model} and \ref{sec:num_res}. These are: (i) neglecting the decay from the Rydberg state during the relevant experimental timescales, (ii) truncating long-range van der Waals or dipolar interactions to nearest-neighbor interactions, (iii) assuming atoms in ground- and Rydberg states to be trapped simultaneously in the same optical lattice, and (iv) neglecting heating due to photons emitted from the $|g\rangle \leftrightarrow |e\rangle$ transition.

(i) The lifetime of Rydberg states is determined by both radiative decay (spontaneous emission) to lower-lying electronic energy levels and blackbody radiation which induces transitions to higher- and lower-lying neighbouring Rydberg states~\cite{rev:rydsaff,rev:rydgall}. While for low-lying electronic levels spontaneous decay to the ground-state dominates, for Rydberg states the interaction with thermally occupied infrared photons, which is the main contribution in the overall lifetime,  leads to strong mechanical effects~\cite{praglaetzle}. In general, (alkali-metal) Rydberg states have, compared to low-lying excited states, exceedingly long lifetimes $\propto n^3$ ($\propto n^5$) for low (high) angular momentum states. For $n\simeq80$, these can be of the order of hundreds of $\mu$s~\cite{beterov}. In our setup, the rate of decay from the Rydberg state $\gamma_r$ can be neglected when
\begin{equation} \label{sdr}
\gamma_r \ll \frac{\Omega_e^2\Omega_r^2 \gamma_e}{4\Delta_r^2\gamma_e^2 + (\Omega_e^2-4\Delta_r^2)^2}
\end{equation}
holds~\cite{rev}, implying that stimulated emission processes from $|r\rangle$ have to be much larger than spontaneous ones, which is typically the case for Rydberg states with large enough principal quantum numbers $n$ and $\Delta_r = 0$. For $\Delta_r \neq 0$, nonetheless, care has to be taken since, contrarily to the case of the general requirement for well-defined quantum jumps to occur~\cite{rev}
\begin{equation} \label{qj}
\Omega_r^2 \ll \frac{4\Delta_r^2\gamma_e^2 + (\Omega_e^2-4\Delta_r^2)^2}{\gamma_e^2 + 4\Delta_r^2},
\end{equation}
the condition in Eq.~(\ref{sdr}) becomes \textit{stricter} with respect to the resonant case $\Delta_r=0$. 

(ii) In Fig.~\ref{fig:cinconf}(b) we check, for a certain initial state and no random detunings, the reliability of the approximation of truncating Rydberg interactions to nearest-neighbour. While for van der Waals  potentials there is no significative quantitative disagreement, dipolar interactions result to be detrimental, despite the qualitative slow growth of the number of Rydberg excitations is recovered in the long time limit. Nevertheless, Rydberg atoms at distances of a few $\mu$m, typical of large spacing optical lattices~\cite{nelson,kuzmich,Browaeys,Schlosser,Hannaford, Spreeuw}, are known to interact with van der Waals  potentials~\cite{rev:rydsaff}, and our approximation is thus not only qualitatively but also quantitatively justified.

(iii) It is important, in our setup, that atoms are trapped in the same optical lattice during both \textit{bright} and \textit{dark} periods. While ground-state atoms can be trapped using different state-of-the-art techniques, it has been suggested to optically trap Rydberg atoms via ponderomotive potentials from spatially modulated light fields~\cite{ponderomotive} or in dc-electric or magnetic fields~\cite{merkt}. In general, electronic ground-states and Rydberg states will experience different optical trapping potentials due to their different polarizabilities. A promising route to trap ground-state atoms and Rydberg-atoms in the same lattice is also to use alkaline-earth atoms as suggested in Ref.~\cite{alkali}.

(iv) Heating due to photons emitted from the strong transition can be considerably reduced adding a finite detuning from $|e\rangle$, $\Delta_e \neq 0$, in order to Doppler-cool atoms with the same fluorescence laser. Results of quantum trajectory simulations (not shown) indicate that the typical slowed relaxation of atoms is qualitatively unaltered, although larger trajectory times with respect to the case  $\Delta_e = 0$ are needed in order to observe the convergence of $r(t)$ to its steady state value $r$.

\section{Conclusions} \label{sec:conc}

We have shown that the non-equilibrium dynamics of Rydberg excitations in three-level atoms can be tuned from the classical to the quantum regime with the laser driving to a short-lived intermediate excited state.  
The anti-blockade condition, where the nearest-neighbor interaction among Rydberg levels equals the single atom laser detuning from the Rydberg state, implements a form of kinetic constraint. This phenomenon resembles the assisted processes of creation and destruction of defects in the One-Spin Facilitated Model. The absence of merging
processes in our system prevents the thermalization to the ground-state of the One-Spin Facilitated Model, despite the concentration of Rydberg excitations can be, in the long time limit, stabilized to some steady state value for sufficiently large interactions and detunings from the Rydberg state.    
In the intermediate regime of interactions, we observed instead a (tuneable) slow increase of the concentration of Rydberg excitations, caused by the appearance of processes which are not present in the One-Spin Facilitated Model. In this regime, an additional slow relaxation timescale emerges.

In the coherent limit, the dynamics of the resulting interacting two-level system is limited to a subspace of the Hilbert space. This subspace is determined  by the choice of the initial state, whose number of clusters of Rydberg excitations is conserved. However, the size of these clusters changes during the evolution. We derived an effective reduced Hamiltonian and verified excellent agreement with simulations of the exact dynamics of the full Hamiltonian.

Finally, we have shown that, in the rate equation limit, the competition of the internal disorder of glasses with an externally imposed disorder can result in the slowing down but also in the speeding up of the overall relaxation towards the steady state. 
The setup we presented allows for the investigation of the quantum dynamics of Spin Facilitated Models in the presence of random fields.
As an outlook, we plan to study the dynamics of the coherent cluster model in the spirit of many-body localization. 

\ack 
We thank P. Zoller for initial discussions which stimulated this work. We further acknowledge M. Fleischauer, J. P. Garrahan,  I. Lesanovski, B. Olmos, M. Dalmonte, M. Heyl and H. Pichler. 
Numerical simulations were performed using QuTip libraries~\cite{qutip}. 
This work is supported by the Austrian Science Fund (FWF): P-25454 N27, the SFB FoQuS (FWF Project No. F4016-N23), the European Research Council (ERC) Synergy Grant
UQUAM, the Austrian Ministry of Science BMWF as part of the UniInfrastrukturprogramm of the Focal Point Scientific Computing at the University of Innsbruck. 
M.M. acknowledges the EU Marie-Curie Program ITN COHERENCE FP7-PEOPLE-2010-ITN-265031 for financial support. 

\newpage

\appendix

\section{Review of the properties of V-shaped three-level atom} \label{appendix:st}
In this Appendix we review properties of a single atom in the V-configuration and describe both quantitatively and qualitatively the physical mechanism of quantum jumps in such a setup.

Let us imagine to measure photons emitted by the three-level atom described by the Hamiltonian in Eq.~(\ref{ham_sa}) of the main text with a photodetector, which we assume to be $100 \%$ efficient. The theory of continuos measurement~\cite{rev} ensures that, under continuous monitoring of the strong transition $|g\rangle \leftrightarrow |e\rangle$ with our photodetector and during time intervals where no photons are detected, the system wave function $|\psi(t) \rangle$ evolves according to the Schr\"{o}dinger equation $i \frac{d}{dt} |\psi(t)\rangle  =  H_{\mathrm{eff}} |\psi(t)\rangle$, with 
\begin{equation} \label{heff}
H_{\mathrm{eff}} = H - \frac{i \gamma_e}{2} c^{\dagger}c = H - \frac{i \gamma_e}{2} |e \rangle \langle e|
\end{equation}
an effective non-Hermitian Hamiltonian and $H$ given by Eq.~(\ref{ham_sa}) in the main text.
The general solution for the system wave function is  

\begin{equation} \label{genwf}
|\psi(t)\rangle = \sum_{n=1}^3 c_n e^{-i\lambda_n t} |u_n\rangle,
\end{equation} 
where $\lambda_n$  and $|u_n\rangle$ are the eigenvalues and associated eigenvectors of $H_{\mathrm{eff}} $, respectively~\cite{lee}. The coefficients $c_n$ can be determined from the initial condition, e.g. $|\psi (0)\rangle  = |g \rangle$. 
In the limit $\Omega_r \ll \Omega_e, \gamma_e, \Delta_r$, $\lambda_n$ and $|u_n \rangle$ can be computed in perturbation theory.

\begin{verse}
\subsection{Brillouin-Wigner perturbation theory calculation of eigenvalues and eigenvectors of $H_{\mathrm{eff}}$} \label{appendix:pt}

In the following, we summarize the steps for the calculation of $\lambda_n$ and $|u_n\rangle$ of $H_{\mathrm{eff}}$ in Eq.~(\ref{heff}) within a Brillouin-Wigner perturbation theory approach. In particular, we focus on the calculation of $\lambda_e$ and $|u_3\rangle$ (the other eigenvalues and eigenvectors can be  straightforardly computed following the same procedure).
In the  $\{ |g \rangle, |e \rangle, |r \rangle\}$ basis, $H_{\mathrm{eff}}$ has the following matrix form  
\begin{equation} \label{heffapp}
H_{\mathrm{eff}} =
\left(\begin{array}{ccc} 0 & \frac{\Omega_e}{2} & \frac{\Omega_r}{2} \\ \frac{\Omega_e}{2} & -\frac{i\gamma_e}{2} & 0 \\ \frac{\Omega_r}{2} & 0 & -\Delta_r \end{array}\right).
\end{equation}  
Assuming $\Omega_r \ll \Omega_e, \gamma_e, \Delta_r$, one can split Eq.~(\ref{heffapp}) in 
\begin{equation}
H_0 =
\left(\begin{array}{ccc}
                        0 & \frac{\Omega_e}{2} & 0 \\
                        \frac{\Omega_e}{2} & -\frac{i\gamma_e}{2} & 0 \\
                        0 & 0 & -\Delta_r
\end{array}\right) \hspace{0.05cm} \mbox{and}  \hspace{0.1cm}
H_1 =
\left(\begin{array}{ccc}
                        0 & 0 & \frac{\Omega_r}{2} \\
                        0 & 0 & 0 \\
                        \frac{\Omega_r}{2} & 0 & 0
\end{array}\right),
\end{equation}  
such that $H_\mathrm{eff}=H_0+H_1$, where $H_0$ is the unperturbed Hamiltonian and $H_1$ is the perturbation.   
Let us introduce a unitary transformation $U = e^{\frac{1}{2} \arctan{ \big \{  \frac{2i  \Omega_e}{\gamma_e} (|e\rangle \langle g | - |g\rangle \langle e|}) \big \}}$ which diagonalizes  $H_0$, 
\begin{equation} \label{h0}
\tilde{H}_0  = UH_0 U^{\dagger} = \left(\begin{array}{ccc} 
                         E_m & 0& 0 \\
                        0& E_p & 0 \\
                       0 & 0 & -\Delta_r
\end{array}\right),
\end{equation}
where $U^{\dagger}$ is the Hermitian conjugate of $U$, $E_m = \frac{i}{4} \left( -\gamma_e+\sqrt{\gamma_e^2-4\Omega_e^2} \right)$ and $E_p =  -\frac{i}{4} \left( \gamma_e+\sqrt{\gamma_e^2-4\Omega_e^2} \right)$.
Applying the same transformation on $H_1$, $\tilde{H}_1 = UH_1U^{\dagger}$, we get 

\begin{equation} \label{h1}
\tilde{H}_1 =
\left(\begin{array}{ccc}                
                       0 & 0& \frac{\Omega_c}{2}\\
                        0& 0 & \frac{i\Omega_s}{2}\\
                       \frac{\Omega_c}{2} & \frac{i\Omega_s}{2} & 0
\end{array}\right),
\end{equation}   
where $\Omega_c = \Omega_r \cosh{ \{ \frac{1}{2} \arctanh{ \frac{2 \Omega_e}{\gamma_e} } \} }$ and $\Omega_s = \Omega_r \, \sinh{ \, \{ \frac{1}{2} \arctanh{ \frac{2 \Omega_e}{\gamma_e} } \} }$.
The unitary transformation $U$ defines a change of basis from $\{ |g\rangle, |e\rangle, |r\rangle \}$ to, say,  $\{ |-\rangle, |+\rangle, |r\rangle \}$

\begin{equation} \label{transf}
\cases{|-\rangle =  \frac{\Omega_c}{\Omega_r} |g\rangle -i\frac{\Omega_s}{\Omega_r} |e\rangle & \\ |+ \rangle = i\frac{\Omega_s}{\Omega_r} |g\rangle + \frac{\Omega_c}{\Omega_r} |e\rangle & \\ |r \rangle  =  |r \rangle.}
\end{equation}
We now define two operators $\mathcal{P}$ and $\mathcal{Q}$ which project
onto $|r \rangle$ and its complementary subspace, respectively. Since $U$ does not affect the $|r\rangle$ subspace [see Eq.~(\ref{transf})], $\mathcal{P}$ has the same matrix form in either the $\{ |g\rangle, |e\rangle, |r\rangle \}$ or the $\{ |-\rangle, |+\rangle, |r\rangle \}$ basis
\begin{equation} 
\mathcal{P} =
\left(\begin{array}{ccc}                
                        0 & 0 & 0 \\
                        0 & 0 & 0 \\
                        0 & 0  & 1
\end{array}\right),
\end{equation}
while $\mathcal{Q}$ can be easily obtained from the orthogonality property of projectors, i.e. $\mathcal{P} + \mathcal{Q} = \mathds{1}$, where $\mathds{1}$ is the identity operator. 
The first- ($E_1$) and second-order ($E_2$) corrections to the eigenvalue $E_0 = [\mathcal{P}H_0\mathcal{P}]_{3,3} =  [\mathcal{P}\tilde{H}_0\mathcal{P}]_{3,3}  = -\Delta_r$ of the transformed unperturbed ($\tilde{H}_0$) Hamiltonian are

\begin{equation}
E_1  = [\mathcal{P}H_1\mathcal{P}]_{3,3} = [\mathcal{P}\tilde{H}_1\mathcal{P}]_{3,3} = 0
\end{equation}
and

\begin{equation}
E_2 = [\mathcal{P}\tilde{H}_1\mathcal{Q}\mathcal{R}_0\mathcal{Q}\tilde{H}_1\mathcal{P}]_{3,3} = -\frac{(\gamma_e + 2i\Delta_r)\Omega_r^2}{4\Delta_r(\gamma_e + 2i\Delta_r)-2i\Omega_e^2}.
\end{equation}
Here, $[A]_{i,j}$ indicates the $i$-th row and $j$-th column of the matrix $A$ and $\mathcal{R}_0 = (E_0 \mathds{1} - \mathcal{Q} H_0 \mathcal{Q})^{-1}$ is the so-called resolvent.
Up to second order in $\Omega_r$, $\lambda_3$ can be approximated with

\begin{eqnarray} \label{lambda3}
\lambda_3  &= \sum_{n=0}^{\infty} E_n \simeq E_0+E_1+E_2 =\nonumber \\ & =  - \Delta_r - \frac{\Omega_r^2(i\gamma_e - 2\Delta_r)}{8\Delta_r^2 - 2\Omega_e^2-4i\gamma_e\Delta_r}.
\end{eqnarray}
We can now calculate the first-order correction to $|r \rangle$ originated by the other two eigenstates of  $\tilde{H}_0$, $|- \rangle$ and $|+ \rangle$, both mixed to $|r \rangle$ by the transformed perturbation Hamiltonian $\tilde{H}_1$. For example, the eigenvector $|u_3\rangle$ associated to $\lambda_3$, is:
\begin{eqnarray} \label{eigvecheff}
|u_3 \rangle  &= | r \rangle + \frac{\langle - | \tilde{H}_1 | r \rangle} {-\Delta_r - E_m} + \frac{\langle + | \tilde{H}_1 | r \rangle} {-\Delta_r - E_p}  = \nonumber  \\ &= | r \rangle -\frac{\Omega_r(i\gamma_e-2\Delta_r)}{2i\gamma_e\Delta_r-4\Delta_r^2+\Omega_e^2}| g \rangle + \nonumber \\ &-\frac{\Omega_e\Omega_r}{2i\gamma_e\Delta_r-4\Delta_r^2+\Omega_e^2}| e \rangle.
\end{eqnarray}
As mentioned above, the coefficients $c_n$ can be calculated from the initial condition. For example, setting $|\psi(0)\rangle = |g\rangle$, we get

\begin{equation}
c_3 = \frac{\Omega_r(i\gamma_e-2\Delta_r)}{4\Delta_r^2-\Omega_e^2-2i\gamma_e\Delta_r}. 
\end{equation}

\end{verse}

\vspace{1.2cm}

The eigenvalues $\lambda_n$ are complex numbers all with negative imaginary parts $\mathrm{Im}\{\lambda_n\}$ [see, e.g., Eq.~(\ref{lambda3})]. As a consequence, $|\psi(t)\rangle$ does not conserve the norm during the evolution.  For $\Omega_r \ll \Omega_e^2/\gamma_e$, 
one of the eigenvalues of $H_{\mathrm{eff}}$, $\lambda_3$ in Eq.~(\ref{lambda3}), has a much less negative imaginary part with respect to the other two, $|\mathrm{Im}\{\lambda_3\}| \ll |\mathrm{Im}\{\lambda_{1}\}|, |\mathrm{Im}\{\lambda_{2}\}|$~\cite{lee}. This determines long tails in the delay function

\begin{equation} \label{d0t}
D_0(t) =  \sum_{n=1}^3 |c_n|^2 e^{2 t \mathrm{Im}\{\lambda_n\}},
\end{equation} 
defined as the conditional probability that no photon has been detected until time $t$, given a photon count has occurred at $t=0$.
Note that  $D_0(t)$  is exactly the norm of $|\psi(t) \rangle$, and its monotonous decay is dominated, in the long time limit, by $\mathrm{Im}\{\lambda_{3}\}$. Thus $\mathrm{Im}\{\lambda_{3}\}$ is responsible of the existence of a small but finite probability of a long time window with no photon counts. When such a long interval of no photon counts occurs, the atom is prepared in the Rydberg state and $|\psi(t) \rangle$ is dominated by $|u_3\rangle$, which in turn can be expressed as $|r \rangle$ plus a perturbatively small admixture of $|g \rangle$ and $|e \rangle$ (see~\ref{appendix:pt}).  

The wavefunction of trajectories, or system realizations, $|\tilde{\psi}_{\alpha}(t)\rangle$ (where $\alpha$ labels each of the $N_{\mathrm{traj}}$ stochastically independent trajectories), can be simulated following the procedure below~\cite{zol_qjumps}: 

\begin{enumerate}
\item Prepare the system in a certain (normalized) state $|\tilde{\psi}(t_0)\rangle_{\alpha}$ at the initial time $t_0$.
\item Compute the probability of the occurrence of a quantum jump at $t_0$ from $p_{\alpha}(t_0) = \gamma_e \langle \tilde{\psi}(t_0) |c^{\dagger} c|\tilde{\psi}(t_0) \rangle_{\alpha}\, dt \ll 1$, where $dt$ is an ideally infinitesimal time step. The corresponding probability of no quantum jump is $1-p_{\alpha}(t_0) \approx 1$. 
\item According to these probabilities, randomly choose a no jump/jump event.\item In case of  no jump, propagate $|\tilde{\psi}(t_0)\rangle_{\alpha}$ with $H_{\mathrm{eff}}$ to obtain the normalised wavefunction $|\tilde{\psi}(t_0+dt)\rangle_{\alpha} = e^{-i H_{\mathrm{eff}} dt}|\tilde{\psi}(t_0)\rangle_{\alpha}/||...|| = (1-iH_{\mathrm{eff}}dt)|\tilde{\psi}(t_0)\rangle_{\alpha}/||...||$. Otherwise, project the wave function into the ground-state, $|\tilde{\psi}(t_0+dt)\rangle_{\alpha} = \sqrt{\gamma_e} \, c |\tilde{\psi}(t_0)\rangle_{\alpha} / ||...|| = |g \rangle$.
\item Repeat the procedure from (i) to (iv), until the desired final trajectory time $t_{\mathrm{max}}$ is reached.
\end{enumerate}
 
\begin{figure}[h!]
\centerline{\includegraphics[width=0.8\textwidth]{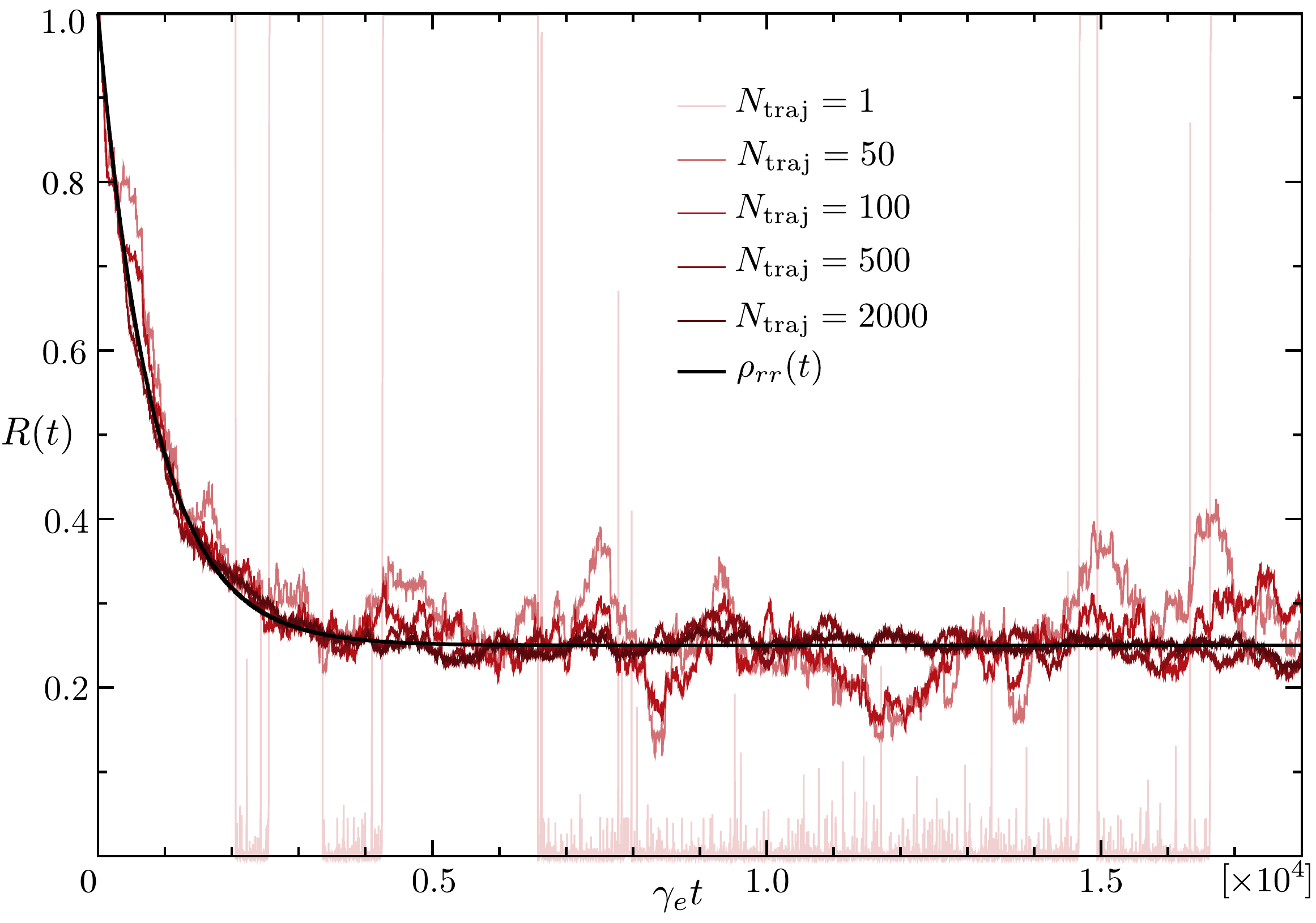}}
\caption{Rydberg population $R(t)$ averaged over $N_{\mathrm{traj}}$ trajectories compared to the exact evolution $\rho_{rr}(t)$ governed by Eq.~(\ref{me}). Parameters are: $\Omega_e = \gamma_e$ and $\Omega_r = 0.03 \gamma_e$. Increasing $N_{\mathrm{traj}}$, $R(t)$  converges to $ \rho_{rr}(t)$, as expected.}
 \label{fig:compmcme}
\end{figure}

A typical experimental sequence of photon detections looks like an alternation of \textit{bright} and \textit{dark} periods. During a \textit{bright} period, labelled by $0$, the atomic population is rapidly cycled between $|g\rangle$ and $|e\rangle$ and the time between two successive photon detections is of the order of the lifetime of $|e\rangle$, $\gamma_e^{-1}$. In contrast, during a \textit{dark} period, labelled with $1$, the atomic population is shelved in $|r\rangle$ and no photons are detected.
An approximation of the exact time evolution of $\rho(t)$ governed by Eq.~(\ref{me}) in the main text can be obtained simulating $N_{\mathrm{traj}}$ trajectories and finally averaging over them, i.e.
\begin{equation}\label{denmat}
\rho(t) = \lim_{N_{\mathrm{traj}} \rightarrow \infty} \frac{1}{N_{\mathrm{traj}}} \sum_{\alpha=1}^{N_{\mathrm{traj}}} |\tilde{\psi}(t) \rangle_{\alpha} \langle \tilde{\psi}(t) |.
\end{equation}
In the following, the observable we are interested in is the Rydberg population, that is, the expectation value of the Rydberg projector $|r\rangle \langle r|$ in $|\tilde{\psi}(t) \rangle_{\alpha}$,

\begin{equation} \label{singler}
r_{\alpha}(t) =  ||\langle r| \tilde{\psi} (t) \rangle_{\alpha}||^2.
\end{equation}
Averaging $r_{\alpha}(t)$ over $N_{\mathrm{traj}}$ leads to

\begin{equation} \label{Ryd}
R(t) = \frac{1}{N_{\mathrm{traj}}} \sum_{\alpha = 1}^{N_{\mathrm{traj}}} r_\alpha(t).
\end{equation}
In analogy to Eq.~(\ref{denmat}), $\lim_{N_{\mathrm{traj}}\rightarrow \infty} R(t) = \rho_{rr}(t)$, where $\rho_{rr}(t) \equiv \langle r | \rho (t) | r \rangle$ is the Rydberg population whose evolution is governed by the exact master equation in Eq.~(\ref{me}). 
Fig.~\ref{fig:compmcme} shows that $R(t)$ converges to $\rho_{rr}(t)$ by increasing $N_{\mathrm{traj}}$.
In the following we illustrate three independent methods for the estimation of the average duration of \textit{bright} and {\it dark} periods $T_0$ and $T_1$, respectively, or equivalently, of the transition probabilities (rates) from $0$ to $1$, $\Gamma^{0 \rightarrow 1} \equiv T_0^{-1}$, and from $1$ to $0$, $\Gamma^{1 \rightarrow 0} \equiv T_1^{-1}$.
In Ref.~\cite{tbtd}, these rates are found to be  

\begin{equation}
\Gamma^{0 \rightarrow 1} = \frac{d \rho_{rr}(t)}{d t} \Bigg |_{t = 0},
\end{equation}
together with $\rho_{rr}(0)=0$, and

\begin{equation}
\Gamma^{1 \rightarrow 0}  = - \frac{d \rho_{rr}(t)}{d t} \Bigg |_{t = 0},
\end{equation}
together with $\rho_{rr}(0)=1$. 
Rates can thus be evaluated by linearization of the short-time evolution of $\rho_{rr}(t)$ with the proper initial condition (e.g., from Fig.~\ref{fig:compmcme}, we can estimate $\Gamma^{1 \rightarrow 0}$).

For comparison, we show the analytical derivation of the rates in Eqs.~(\ref{sprateup}) and~(\ref{sprate}) of the main text. In the long time limit $D_0(t) \sim p \, e^{- t/T_1}$, where $p$ is the (small) probability to be in a \textit{dark} period~\cite{cohen_dal,rev}. Comparing it with Eq.~(\ref{d0t}) one gets

\begin{equation} \label{sprate1}
\Gamma^{1 \rightarrow 0} (\Delta_r) = -2\, \mathrm{Im}\{\lambda_3\} = \frac{\gamma_e \Omega_e^2 \Omega_r^2}{16 \Delta_r^4 + 4 \Delta_r^2 (\gamma_e^2 - 2\Omega_e^2) + \Omega_e^4},
\end{equation}
and 

\begin{equation}
p = |c_3^2| = \frac{\Omega_r^2(\gamma_e^2 + 4\Delta_r^2)}{16 \Delta_r^4 + 4 \Delta_r^2 (\gamma_e^2 - 2\Omega_e^2) + \Omega_e^4},
\end{equation}
where both results are obtained substituting $\lambda_3$ and $c_3$ calculated in~\ref{appendix:pt}. In order to compute $\Gamma^{0 \rightarrow 1}(\Delta_r)$, we note that, during a \textit{bright} period, $|r\rangle$ is almost never populated and the effective emission rate  will be that of the (resonantly driven) two-level system $|g\rangle \leftrightarrow |e\rangle$, i.e. $\gamma_e \Omega_e^2/(\gamma_e^2 + 2\Omega_e^2)$. Then, $\Gamma^{0 \rightarrow 1}(\Delta_r)$  is simply the two-level emission rate multiplied by $p$~\cite{lee},

\begin{equation} \label{sprateup1}
\Gamma^{0 \rightarrow 1} (\Delta_r) = \frac{\gamma_e \Omega_e^2\Omega_r^2(\gamma_e^2 + 4\Delta_r^2)}{(\gamma_e^2 + 2\Omega_e^2)[16 \Delta_r^4 + 4 \Delta_r^2 (\gamma_e^2 - 2\Omega_e^2) + \Omega_e^4]}.
\end{equation}
Eqs.~(\ref{sprate1}) and~(\ref{sprateup1}) are exactly Eqs.~(\ref{sprateup}) and Eqs.~(\ref{sprate}) in the main text, respectively.
If $\Delta_r =0$ they simplify to 

\begin{equation} \label{sprateres}
\Gamma^{1 \rightarrow 0} (0) \equiv \Gamma^{1 \rightarrow 0} = \frac{\gamma_e \Omega_r^2}{\Omega_e^2} \hspace{0.6cm} \mbox{and} \hspace{0.6cm} \Gamma^{0 \rightarrow 1} (0) \equiv \Gamma^{0 \rightarrow 1} = \frac{\gamma_e^3 \Omega_r^2}{\Omega_e^2 (\gamma_e^2 + 2\Omega_e^2)}.
\end{equation}

The third independent method relies on the direct numerical calculation of the average over trajectories of the time-averaged length of \textit{dark} and \textit{bright} periods in each trajectory.
Results for $\Gamma^{1 \rightarrow 0}$, obtained with the three different methods, are compared in Tab.~\ref{tab:rates}:  

\begin{table}[h!]
\centering
 \begin{tabular}{||c | c | c | c||} 
 \hline
& $ME$ & $PT$ &  $QT$ \\ [0.5ex] 
 \hline
$\Gamma^{1 \rightarrow 0}/\gamma_e$ &  $9.02 \cdot 10^{-4}$ &   $9.00 \cdot 10^{-4}$  &  $9.13 \cdot 10^{-4}$   \\  [1ex] 
 \hline
 \end{tabular}
 \caption{Estimates of the resonant [$\Delta_r = 0$] rate $\Gamma^{1 \rightarrow 0}$ (in units of $\gamma_e$), according to: the short time slope of $\rho_{rr}(t)$, obtained as the exact solution of the master equation  ($ME$) [Eq.~(\ref{me}) of the main text], perturbation theory ($PT$) [Eq.~(\ref{sprateres})] and quantum trajectory simulations ($QT$). In all three methods, parameters are: $\Omega_e = \gamma_e$ and $\Omega_r = 0.03 \gamma_e$. Within the $QT$ method, simulations of $N_{\mathrm{traj}} = 6000$ trajectories up to times  $\gamma_e t= 10^6$ have been performed.}
 \label{tab:rates}
\end{table}

\section{Steady state Rydberg population} \label{appendix:ss}

The steady state solution of the single atom master equation can be obtained setting $\dot{\rho} (t) = 0$ in Eq.~(\ref{me}) of the main text. This is equivalent to calculate the kernel of the matrix of the corresponding optical Bloch equations 

\begin{equation} \label{obe3lev}
\frac{\partial}{\partial t}  \rho =
\left(\begin{array}{ccccccccc}               
                        -\frac{\gamma_e}{2} & 0 & - \frac{i\Omega_e}{2} &   \frac{i\Omega_e}{2} & 0 & 0 & 0 & - \frac{i\Omega_r}{2} & 0\\
                         0 & -\frac{\gamma_e}{2} &  \frac{i\Omega_e}{2} &  - \frac{i\Omega_e}{2} & 0 & 0 & 0 & 0 &  \frac{i\Omega_r}{2}\\
                         - \frac{i\Omega_e}{2} &  \frac{i\Omega_e}{2} & - \gamma_e & 0 & 0 & 0 & 0 & 0 & 0\\
                          \frac{i\Omega_e}{2} & - \frac{i\Omega_e}{2} &  \gamma_e & 0 &  \frac{i\Omega_r}{2} & -  \frac{i\Omega_r}{2} & 0 & 0 & 0\\
                         0 & 0 & 0 & \frac{i\Omega_r}{2} &0 & 0 & -  \frac{i\Omega_r}{2} & 0 & -  \frac{i\Omega_e}{2} \\
                         0 & 0 & 0 & - \frac{i\Omega_r}{2} & 0 & 0 &  \frac{i\Omega_r}{2} & \frac{i\Omega_e}{2}  & 0  \\
                         0 & 0 & 0 &0 & -  \frac{i\Omega_r}{2} &   \frac{i\Omega_r}{2}& 0 & 0 & 0  \\
                         - \frac{i\Omega_r}{2} & 0 & 0 &0 & 0 &  \frac{i\Omega_e}{2}& 0 &  - \frac{\gamma_e}{2} & 0  \\
			0 & \frac{i\Omega_r}{2} & 0 &0 & - \frac{i\Omega_e}{2} & 0& 0 & 0 &  - \frac{\gamma_e}{2}  \\
\end{array}\right) \rho,
\end{equation}
where $\rho = ( \rho_{eg}, \rho_{ge}, \rho_{ee}, \rho_{gg},  \rho_{rg}, \rho_{gr}, \rho_{rr}, \rho_{er},  \rho_{re} )^T$,   $\rho_{\alpha \beta} = \langle \alpha | \rho | \beta \rangle$  and $ \alpha, \beta = g,e,r$. (Here, we have suppressed the explicit time dependence of all terms, i.e. $\rho_{\alpha\beta} \equiv \rho_{\alpha\beta}(t)$, in order to improve the readability.)
Since the matrix in Eq.~(\ref{obe3lev}) is singular, we have an infinite number of steady state solutions parametrized by the (non-normalized) vector

\begin{equation}
\rho(t \rightarrow \infty) =
                         \left ( i \frac{\gamma_e}{\Omega_r}, 
                         - i \frac{\gamma_e}{\Omega_r},
                         \frac{\Omega_e}{\Omega_r}, 
                         \frac{\gamma_e^2+\Omega_e^2+\Omega_r^2}{\Omega_e\Omega_r} ,
                         0,  
                         0 ,
                         \frac{\gamma_e^2+\Omega_r^2}{\Omega_e\Omega_r} ,
                         1,
                         1 \right )^T.
\end{equation}
Using the property of trace conservation of $\rho(t)$,  $\,$Tr$\{ \rho(t) \} = \rho_{gg}(t) + \rho_{ee}(t)  +\rho_{rr}(t)  = 1$, the normalized steady state Rydberg population can finally be written as

\begin{equation} \label{ssrr}
\rho_{rr}(t\rightarrow \infty) = \frac{\gamma_e^2 + \Omega_r^2}{2(\gamma_e^2 + \Omega_e^2 + \Omega_r^2)}.
\end{equation}
Notice that Eq.~(\ref{ssrr}) converges, for $\Omega_r \ll \Omega_e, \gamma_e $, to the equilibrium concentration of Rydberg excitations $d_{\mathrm{eq}}$ [Eq.~(\ref{deq}) in the main text].

\end{document}